\documentclass[prd,twocolumn,amsmath,amssymb,nofootinbib]{revtex4-2}
\usepackage{physics}
\usepackage{lmodern}
\usepackage{xcolor}
\usepackage{url}
\usepackage{hyperref}
\usepackage{cleveref}
\usepackage{graphicx}
\usepackage{subfigure}
\usepackage{comment}
\usepackage{tikz}
\usepackage{tikzlings}
\usetikzlibrary{calc, patterns, decorations, decorations.markings, shapes, positioning, intersections, quotes, angles}
\usetikzlibrary{shapes.misc} % <-- This fixes the 'cross out' error
\usepackage{amsmath}
\usepackage{mathtools}

\usepackage{xcolor}
\usepackage{colortbl}
\definecolor{dark-red}{rgb}{0.50,0.12,0.12} % links
\definecolor{mblue}{rgb}{0.30, 0.45, 0.70}
\definecolor{mred}{rgb}{0.70, 0.20, 0.20}
\definecolor{mgray}{rgb}{0.63, 0.63, 0.63}

\usepackage[utf8]{inputenc}
\usepackage[T1]{fontenc}
\hypersetup{
  colorlinks=true,
  citecolor=magenta,
  linkcolor=blue,
  urlcolor=violet
 }
\DeclareUnicodeCharacter{2212}{\ensuremath{-}}

\newcommand{\beq}{\begin{equation}}
\newcommand{\eeq}{\end{equation}} 

\newcommand{\lc}{\left(}
\newcommand{\rc}{\right)}
\newcommand{\ls}{\left[}
\newcommand{\rs}{\right]} 
\def\d{{\rm d}}

\newcommand{\ic}{\mathrm{i}}
\newcommand{\diff}{\mathrm{d}}

\newdimen\hfuzz
\newdimen\vfuzz

\begin{document}

\title{Imprint of the black hole singularity on thermal two-point functions}

\author{Nima Afkhami-Jeddi$^1$}
%\email{nimaaj@physics.mcgill.ca}

\author{Simon Caron-Huot$^1$}
%\email{schuot@physics.mcgill.ca}

\author{Joydeep Chakravarty$^1$}
%\email{joydeep.chakravarty@mail.mcgill.ca}

\author{Alexander Maloney$^{1,2,3}$}
%\email{alex.maloney@mcgill.ca}

\affiliation{$^1$Department of Physics, McGill University, Montr\'eal, Qu\'ebec, H3A 2T8, Canada}

\affiliation{$^2$Department of Physics, Syracuse University, Syracuse, NY, 13244, USA }

\affiliation{$^3$Institute for Quantum and Information Sciences, Syracuse University Syracuse, NY, 13244, USA}

\begin{abstract}
We consider two-point functions of light fields at finite temperature and large real frequencies in holographic theories.
The thermal system is dual to a single-sided anti-de Sitter black hole. 
We show that the high-frequency expansion obtained from the operator product expansion receives exponentially small nonperturbative corrections, which are controlled by null geodesics bouncing off the black hole singularity in the two-sided eternal black hole geometry. We develop a bulk WKB description of these bouncing geodesics and explain how to calculate reflection coefficients at the singularity.
% We show that, in holographic theories, two-point functions at large real frequencies are controlled by 
% We study the spectral density of four-dimensional $\mathcal{N}=4$ strongly coupled super-Yang-Mills in the large N limit at finite temperature. The thermal system is dual to a single-side AdS$_5$ black hole. At large real frequency, we find that the subleading contribution to the spectral density for light fields is explained by null geodesics reflected off the singularity in the eternal black hole. 
\end{abstract}

\maketitle

\newpage
\section{Introduction}
Within holography \cite{Maldacena:1997re, Gubser:1998bc, Witten-ads-and-holography}, specific field theories at finite temperature are dual to black holes in asymptotically AdS$_5$ spacetime at the corresponding Hawking temperature. This framework relates the physics of  ordinary, albeit strongly interacting, quantum systems to various calculable quantities in the black hole background \cite{Son:2002sd, Policastro:2001yc, Policastro:2002se}.

Perhaps surprisingly, geodesics approaching the singularity in the interior of the black hole %geometry
lead to features in exterior correlators \cite{Fidkowski:2003nf, Festuccia:2005pi, Festuccia:2008zx, Ceplak:2024bja}. This connects the still-mysterious dynamics of probes near the singularity to well-defined exterior quantities.
The features can be revealed by considering heavy bulk fields and analytically continuing correlations between the two sides of the thermofield double \cite{Maldacena:2001kr}.  
The proper time to the singularity can be similarly observed
\cite{Grinberg:2020fdj}.

In this paper we discuss the effect of geodesics that bounce off the singularity in a comparatively simpler setup: single-sided two-point functions with large real frequency and small operator dimension. An example is the spectral density of R-charge currents at zero spatial momentum in $\mathcal{N}=4$ super Yang-Mills, which was observed numerically in \cite{Teaney:2006nc}, and later analytically \cite{Myers:2007we}, to approach its zero-temperature value exponentially fast at high frequencies:
\beq \label{chi}
 \begin{split}
{\rm Im}\, G^{xx}_{\rm ret}(\omega,q=0) &= {\pi \omega^2 (1 - e^{- {\beta \omega }}) \over (1 - e^{-{\beta \omega \over 2}(1-\ic)})(1 - e^{-{\beta \omega \over 2}(1+\ic)})} \\
&\hspace{-15mm}= \pi \omega^2\lc  1 + \sum_{n=1}^\infty (e^{-{n\beta \omega \over 2}(1-\ic)} + e^{-{n\beta \omega \over 2}(1+\ic)}) \rc.
\end{split}
 \eeq
The second line shows the high-frequency expansion, with the leading term $\pi \omega^2$ being the zero-temperature result. We will show how the two terms with $n=1$ can be attributed
%are produced
to a bouncing geodesic and its time reflection, with higher $n$ corresponding to multiple reflections. 

More generally, we will study the high-frequency behavior of the retarded function dual to a scalar field in the AdS$_5$ black hole geometry, where exact solutions are not available; the result will involve a transseries.  We will develop a bulk description using the WKB approximation near null geodesics 
%(see also \cite{Caron-Huot:2025hmk, Caron-Huot:2025she}) 
and a novel reflection coefficient capturing the effect of the singularity.

By definition, the retarded response function and its associated spectral density can be measured and/or calculated using only the geometry in the exterior of the black hole.  The relative simplicity of this observable is one of its key attractive features: in essence one has to measure thermal conductivity as a function of frequency and subtract off the vacuum contribution.

The fact that this observable receives contribution from geodesics that go through the \emph{interior} of the black hole may seem surprising.  In essence, this happens because the highly oscillatory Fourier transform at high frequency will be dominated by complex times, thus revealing complex solutions to the bulk geodesic equations.  The fact that the geodesics appear to explore the interior does not contradict causality, since, as will be clear from the calculation, the bouncing geodesics really only explore an analytic continuation of the exterior geometry.

We can thus already anticipate that our description of the bouncing geodesics will not have any obvious relevance to the experience of an infalling observer, nor will shed light on fine-grained aspects of the information carried by black holes.  Nonetheless, the singularity in the (analytically continued) geometry leaves well-defined imprints in exterior observables, which we will specifically discuss for $\mathcal{N} =4$ super Yang-Mills in four dimensions.

\section{Bulk setup and WKB phase} 

We consider the momentum space retarded thermal Green function in the CFT$_d$:
\beq \label{Gret def}
\hspace{-2mm}
G_{\rm ret}(\omega,q) = \ic\!\!\int\!\d^{d-1}x\int_0^\infty\! \d t \, e^{\ic \omega t-iq{\cdot}x}   \langle \ls \mathcal{O}(t,x),   \mathcal{O}(0) \rs\rangle_{\beta} .
\eeq
Let us briefly review its calculation in a holographic theory. The bulk dual of the thermal state is the AdS$_{d+1}$ planar black hole metric
\beq \label{meth}
{\rm ds}^2 = -r^2f(r) { {\rm d}t}^2 + \frac{{\rm d}r^2}{r^2f(r)} + r^2 {\d  \rm x}^2
\eeq 
where we set the AdS radius $\ell=1$; $f(r) = 1 {-} (r_h/r)^d$ with $r_h=\frac{4\pi}{\beta d}$ the horizon radius and ${\d \rm x}^2$ denotes the flat metric over the $(d{-}1)$ spatial directions.
The bulk field dual to the scalar operator $\mathcal{O}$ of dimension $\Delta$ 
is a scalar of mass $m^2=\Delta(\Delta-d)$
satisfying the equation of motion
\begin{equation} \label{eom}
\phi''+\left(\frac{f'}{f}+\frac{d+1}{r}\right)\phi'
+\frac{\omega^2/f-q^2-m^2r^2}{r^4f}\phi=0 ,
\end{equation}
where primes denote derivatives with respect to $r$.
According to \cite{Son:2002sd, Policastro:2001yc, Policastro:2002se}, the retarded function can be obtained by finding
the solution $\overleftarrow{\phi}_{\omega,q}(r)$ that is purely infalling at the horizon, and
dividing the coefficients of the two independent powers of $r$ near the AdS boundary ($r\to\infty$):
\begin{equation}\label{recipe}
    G_{\rm ret}(\omega,q) = C \left.\left(\overleftarrow{\phi}_{\omega,q}|_{r^{-\Delta}}\right)\middle/\left(\overleftarrow{\phi}_{\omega,q}|_{r^{\Delta-d}}\right)\right.
\end{equation}
The normalization $C$ cancels out when dividing by zero-temperature results and will thus be ignored below.

\begin{figure}[t!]
    \centering
\includegraphics[width=0.99\linewidth]{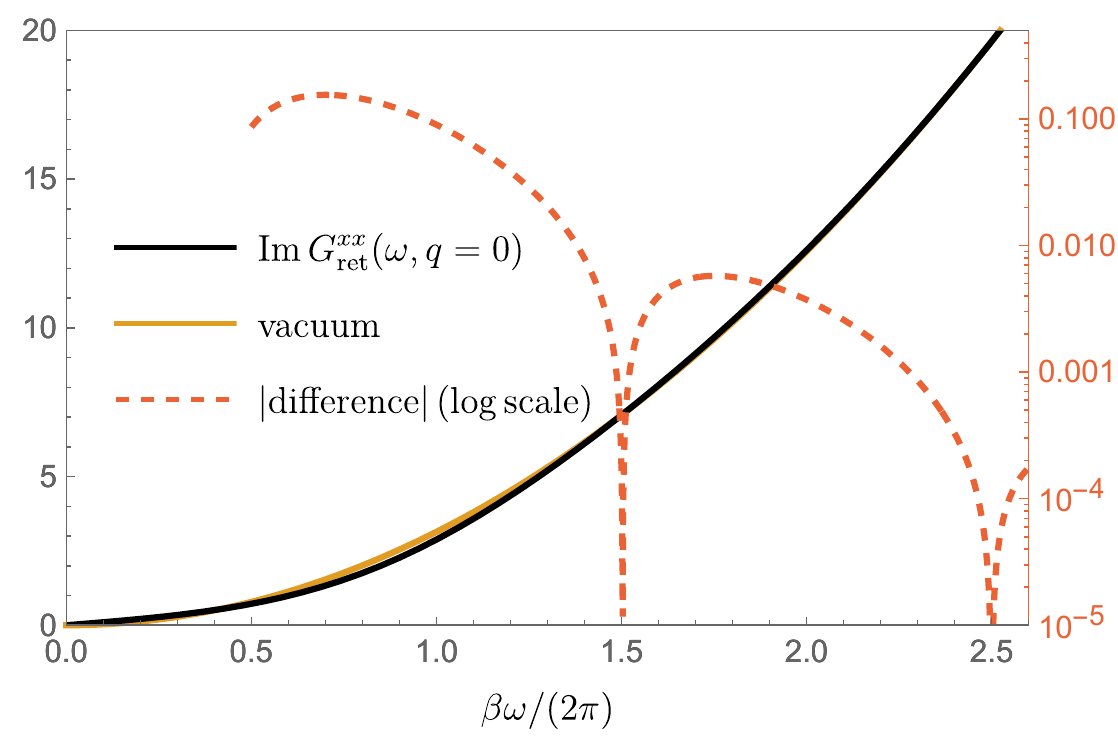}
    \caption{Current spectral density \eqref{chi} at finite and zero temperature. We will focus on the exponentially decaying difference at large frequencies; the dips are caused by destructive interference between the two paths in Fig.~\ref{fig:geod1}.}
    \label{fig:current}
\end{figure}

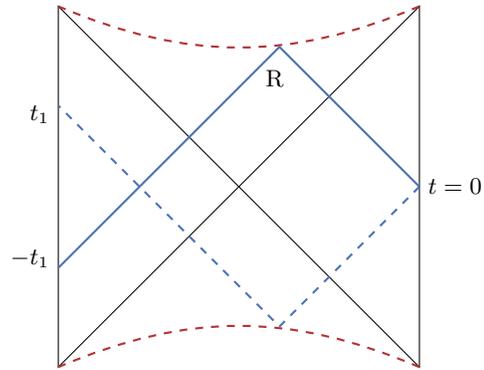
\begin{figure}[t!]
    \centering
    \begin{tikzpicture}[scale=0.60]
\node (I)    at ( 4,0)   {};
\node (II)   at (-4,0)   {};
\path  % Four corners 
  (II) +(90:4)  coordinate[label=90:]  (IItop)
       +(-90:4) coordinate[label=-90:] (IIbot)
       +(0:4)   coordinate                  (IIright)       ;
\draw (IItop) --
      (IIright) -- 
      (IIbot)--cycle;
\path % Four conners of the right diamond (no labels this time)
   (I) +(90:4)  coordinate (Itop)
       +(-90:4) coordinate (Ibot)
       +(180:4) coordinate (Ileft)
       ;
\draw  (Ileft) -- (Itop) -- (Ibot) -- (Ileft) -- cycle;
% Squiggly lines
\draw[dashed, mred, thick] (IItop) to[out=-23,in=-157] (Itop)
      node[midway, above, inner sep=2mm] {};
\draw[decorate,dashed, mred, thick] (IIbot) to[out=+23,in=-203] (Ibot) 
      node[midway, below, inner sep=2mm] {};
%geodesics      
\draw[mblue, thick] (4,0) -- (0.9, 3.1)--(-4,-1.8);
%\draw[mgray!60, thick] (-0.4,4.4) -- (0.9, 3.1);
%\node[right]   at (0.4,3.8)   {$\rm T$};
\node[right]   at (0.4,2.4)   {$\rm R$};
%\draw[mgray, thick, dashed] (-0.4,-4.4) -- (0.9, -3.1);
\draw[mblue, thick, dashed] (4,0) -- (0.9,-3.1) --(-4,1.8) ;
%label points
\draw[fill] ($(Itop)!.5!(Ibot)$)  node[right] {    $t=0$ };
\draw[fill] ($(IItop)!.7!(IIbot)$)  node[left] { $-t_{1}$};
\draw[fill] ($(IItop)!.3!(IIbot)$)  node[left] { $t_1$};
\end{tikzpicture}
    \caption{Nonperturbative contributions $\sim e^{-\frac12\beta\omega+\ic \omega t_1}$ to the retarded function at high frequencies will be explained from a null geodesic that reflect once off future singularity of the eternal black hole.
    Wightman functions and ${\rm Im}\, G_{\rm ret}$ also receive contributions from the time-reversed
  geodesic.} %  bouncing off the past singularity.}
    \label{fig:geod1}
\end{figure}

The large-$\omega$ limit motivates a WKB approximation to the radial solution in the limit $\omega\gg m,q$:
\begin{equation} \label{WKB}
    \phi_{\omega,q}\approx r^{\frac{1-d}{2}} \exp\!\left(\!\ic \omega\!\!\int \!\d t+\mathcal{O}(1/\omega)\right), \quad \d t=\frac{\d r}{r^2f(r)} .
\end{equation}
This was discussed recently for null geodesics in AdS in \cite{Dodelson:2023nnr, Dodelson:2023vrw, Caron-Huot:2025hmk, Caron-Huot:2025she}.
A first hint that null geodesics bouncing off the singularity are relevant to this limit comes from calculating the exponent %boundary time
for the path shown in Fig.~\ref{fig:geod1}, which goes to the singularity and back out to the other side \cite{Amado:2008hw}:
\beq \label{deltatau}
\Delta t = 2\int_{0}^\infty \frac{\d r }{r^2f(r)}  = {\beta \over 2}\lc \cot{\pi \over d} + \ic \rc \xrightarrow[d=4]{} {\beta \over 2}\lc 1+\ic\rc.
\eeq
We see that the factor $e^{\ic \omega\Delta t}$ 
for this path precisely matches the exponentials in \eqref{chi}.
The imaginary part of the travel time originates from the pole of $\d t$ at the black hole horizon, which we avoided with a $-\ic 0$ shift (to be explained below). It produces the half-Boltzmann suppression $e^{-\beta\omega/2}$ visible in \eqref{chi} and represents the fact that the geodesic ends on the left boundary despite the original retarded correlator being defined entirely on the right boundary! The real part of $\Delta t$ is the vertical offset in Fig.~\ref{fig:geod1}. It is nonvanishing for $d>2$, signifying that the Penrose diagram of the AdS black hole is ``not a square'' \cite{Klosch:1995qv, Fidkowski:2003nf}.

The calculation in \eqref{deltatau} is suggestive but incomplete: the WKB approximation breaks down near the singularity and we still need to explain why the trajectory reflects there.
This will be done in section \ref{sec:proposal}.
Before that, we explain on general grounds why the retarded correlator at large \emph{real} frequencies and small $\Delta$ probes the singularity; these kinematics differ from the large mass or large imaginary frequencies previously discussed in the literature.

\section{Operator product expansion}

Key properties of high-frequency two-point functions can be understood using the field theory Operator Product Expansion,
which is most readily stated in terms of the definite-order (Wightman) functions:
\begin{equation}
    G^>(t,x) \equiv \langle  \mathcal{O}(t,x)   \mathcal{O}(0)\rangle_\beta,
\quad G^<(t,x)\equiv \langle  \mathcal{O}(0)   \mathcal{O}(t,x)\rangle_\beta.
\end{equation}
The function $G^>$ coincides with the Euclidean correlator when $t=-\ic\tau$ with $0\!<\!\tau\!<\!\beta$, where it admits the OPE \cite{Iliesiu:2018fao}
\begin{equation} \label{OPE general}
    G^>(-\ic\tau,x)
    = \sum_{\Delta',\ell}
     \frac{C_\ell^{(d/2-1)}(\frac{\tau}{\sqrt{x^2+\tau^2}})}{(\sqrt{x^2+\tau^2})^{2\Delta_{\cal O}-\Delta'}}
   c_{\Delta',\ell}  \langle \mathcal{O}_{\Delta',\ell}\rangle_\beta  
\end{equation}
which runs over the scaling dimension and spin of operators; $c$ are their OPE coefficients and $C_\ell$ is a Gegenbauer polynomial. While the general structure of this expansion follows from the translation and rotational invariance of the thermal state, an important fact is that this sum has a finite radius of convergence. For example, convergence for $\tau$ real and $\sqrt{x^2+\tau^2}<\beta$ was established in \cite{Iliesiu:2018fao}.

The expansion \eqref{OPE general} can thus be analytically continued to real times, and the commutator $G_{\rm ret}=\ic(G^>-G^<)$ can be calculated as the
\emph{discontinuity} of \eqref{OPE general} across the real positive time axis, at least for sufficiently small times.
Which operators contribute to the OPE of $G_{\rm ret}$?

% connected stress-tensor two-point function $\sim N^2$

At large $N$, the OPE of thermal correlators is saturated by products of stress tensors and by ${\cal O}{\cal O}$ double traces, whose scaling dimensions satisfy, respectively, $\Delta=nd$ and $\Delta-\ell=2n+{\cal O}(N^{-2})$ for integer $n\geq 0$ (see \cite{Buric:2025anb, Buric:2025fye} and also \cite{Barrat:2025nvu, Barrat:2025twb}).
Since the contribution of the latter to \eqref{OPE general} is \emph{polynomial}, they cancel out to leading order in the commutator \eqref{Gret def}.
Hence, in a holographic theory, the OPE of the retarded function is saturated at large $N$ by products of stress tensors.
% light operators can be organized into products of single-traces, which in a holographic theory are by definition sparse and consist of protected operators such as the stress tensor, currents or the scalar ${\cal O}$ we are considering, other single-traces acquiring large scaling dimensions \cite{Heemskerk:2009pn}.
% Let us assume for simplicity that the only single-trace which has a nonzero expectation value at leading order in $N$ is the stress tensor, meaning, for example, that all chemical potentials vanish. In terms of canonically normalized operators (whose vacuum two-point functions are of order one), the expectation value of the $k$'th power of the stress tensor is of order $\langle \tilde{T}^k\rangle\sim N^k$, while its 
% OPE coefficient is of order $N^{-k}$. Hence, products of arbitrarily many stress tensors (without derivatives) contribute at leading order.
%The other important operators are double-traces $\sim{\cal O}(\partial^2)^n\partial^{\mu_1}\cdots \partial^{\mu_\ell}{\cal O}$ with angular momentum $\ell$, whose OPE coefficients are of order 1 and whose connected thermal expectation value are also of order 1.
Specializing to zero spatial momentum for notational simplicity (this discussion is easily generalized), the retarded function thus admits an expansion:
\begin{equation} \label{Gret t}
    G_{\rm ret}(t,q=0) = t^{d-1-2\Delta_{\cal O}}
    \sum_{n=0}^\infty a_n \left(\frac{t}{\beta}\right)^{nd}\quad(t>0) .
\end{equation}
This property is explicitly seen in holographic calculations, cf. Appendix \ref{app:expansions}.\footnote{When $2\Delta_{\cal O}+2k=nd$ for some non-negative $k,d$, or when ${\cal O}$ is the stress tensor, the distinction between double-traces and stress tensors 
becomes singular and correlators feature logarithms. Nonetheless, these logarithms cancel when taking the discontinuity and the form \eqref{Gret t} remains valid.

Independently, when $2\Delta_{\cal O}-(n+1)d$ is a nonnegative integer, the $\Gamma$-function in \eqref{Gret omega} develops poles due to the standard logarithmic ultraviolet divergence of the Fourier transform; ${\rm Im}\,G_{\rm ret}(\omega)$ discussed below remains finite and unambiguous.
}

It would be interesting to characterize the convergence and potential singularities of \eqref{Gret t}.
A simple fact is that the commutator only has support for $|x|<t$.
Hence, at small $t$, the spatial Fourier transform runs over a finite $x$ range,
which suggests a finite radius of convergence in a general theory.

Fourier-transforming \eqref{Gret t} term by term gives
\begin{equation} \label{Gret omega}
    G_{\rm ret}(\omega,q=0) \sim (-\ic\omega)^{2\Delta_{\cal O}-d}
\sum_{n=0}^\infty \frac{a_n}{(-\ic \beta\omega)^{nd}} 
\Gamma(d{-}2\Delta_{\cal O}{+}nd).
\end{equation}
We see that the convergent small-time expansion has turned into an asymptotic series in $1/\omega$, due to the factorial growth of the $\Gamma$-function.

The spectral density $2{\rm Im} G_{\rm ret}$ is of particular interest since it determines frequency space Wightman functions by the Kubo-Martin-Schwinger (KMS) conditions:
\begin{equation} \label{kms0}
    G^>(\omega,q)= \frac{2{\rm Im} G_{\rm ret}(\omega,q)}{1-e^{-\beta\omega}} .
\end{equation}
Taking the imaginary part of \eqref{Gret omega} reveals an interesting feature: when $d$ and $2\Delta_{\cal O}$ are both even integers, the series terminate!
As noted previously in \cite{Caron-Huot:2009ypo}, this explains why thermal contributions to the
spectral density of currents and stress tensors in $d=4$ holographic theories decay exponentially, as was already illustrated in Fig.~\ref{fig:current}.%

% Momentum space Wightman functions admit similar expansions, for example
% \begin{equation}
%  G^>(\omega,q=0) = \frac{2 {\rm Im} G_{\rm ret}(t,q=0)}{1-e^{-\beta\omega}}
% = \sum_{n=0}^{\infty} 
% \end{equation}

\begin{figure}[t!]
\begin{tikzpicture}[scale=0.7]
  % Axes
  \draw[ ->] (0, -2.35) -- (0, 3.65) (-4.8,0)--(4.8,0);
 % Poles
      \node at (0, 0) [mred, cross out, draw, thick, inner sep=2pt, minimum size=2pt] {};
      \node at (4,2) [mred, cross out, draw, thick, inner sep=2pt, minimum size=2pt] {};
      \node at (-4,2) [mred, cross out, draw, thick, inner sep=2pt, minimum size=2pt] {};
    \node at (4,-2) [mred, cross out, draw, thick, inner sep=2pt, minimum size=2pt] {};
      \node at (-4,-2) [mred, cross out, draw, thick, inner sep=2pt, minimum size=2pt] {};
%blue contour around (2,2)
 \draw[thick, mblue] (4 -0.2,2) arc[start angle=-180, end angle=0, radius=0.2];
   \draw[thick, mblue] (4 -0.2,3.5) --  (4 -0.2,2) ;
      \draw[thick, mblue, ->] (4 +0.2,2) --  (4 +0.2,3.5) ;
%blue contour about (0,0)
      \draw[thick, mblue, ->] (0 +0.2,0) --  ( +0.2,3.5) ;
\draw[thick, mblue] (0, -0.2) arc[start angle=-90, end angle=0, radius=0.2];
  % Red contour along real axis
  \draw[thick, mred, ->] (0.25,0) -- (4.8,0) ;
  \draw[thick, mred] (0, -0.25) arc[start angle=-90, end angle=0, radius=0.25];
  %branch cuts
 \draw[decorate,decoration=zigzag, mred, thick] (4,2) --(4,3.5);
 \draw[decorate,decoration=zigzag, mred, thick] (-4,-2) --(-4,-2.55);
  \draw[decorate,decoration=zigzag, mred, thick] (4,-2) --(4,-2.55);
  \draw[decorate,decoration=zigzag, mred, thick] (-4,2) --(-4,3.5);
\draw[decorate,decoration=zigzag, mred, thick] (0,0) --(0,3.5);
 % Blue contours
\node at (6,3) {$t$};
 \draw[] (6.2, 2.8) --  (5.8, 2.8) --(5.8, 3.2);
\end{tikzpicture}
\caption{The original real-time contour (in red) for the Fourier transform of $G_{\rm ret}(t)$ can be deformed into a steepest-descent contour along the imaginary axis plus a branch cut starting at $ t = \frac{\beta}{2} (1 + \ic)$. The same cut appears in the Wightman functions $G^>(t)$ continued to the second sheet.}
\label{fig:contour}
\end{figure}
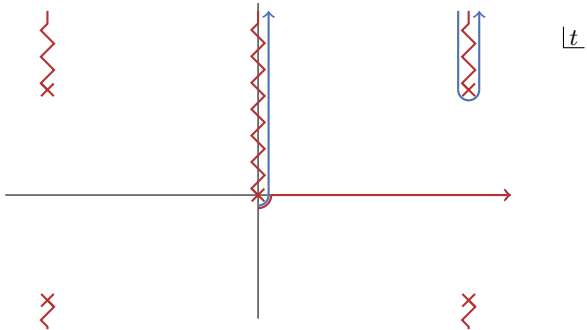

For generic values of $\Delta_{\cal O}$ or $d$, both the retarded and Wightman functions admit nonterminating asymptotic expansions in $1/\omega$.  Borel resumming these expansions is equivalent to analyzing the Fourier integral:
\begin{equation}
    G_{\rm ret}(\omega,q)=\int_0^\infty \d t  \,e^{\ic \omega t} G_{\rm ret}(t,q) .
\end{equation}
A simple steepest-descent analysis of this integral explains why the behavior at large \emph{real} frequencies is controlled by complex times:
for large positive $\omega$, it is advantageous to deform the time contour into the upper-half-plane where the Fourier factor decays.
The deformed contour picks up the ``Borel'' singularities of $G_{\rm ret}$ at complex times, as depicted in Fig.~\ref{fig:contour}. A singularity at $t=t_*$ thus produces a nonperturbative effect $\sim e^{\ic t_*\omega}$ in the $1/\omega$ expansion.

Below we will focus on the bouncing geodesic singularities which appear at ${\rm Im}\,t_*=\beta/2$. This complex time value implies geometrically that the geodesic ends on the left boundary, as shown in Fig.~\ref{fig:geod1},
even though we emphasize that the original correlator involves only the right boundary.  Since $G_{\rm ret}=\ic(G^>-G^<)$ where $G^<$ is analytic in the strip $0<{\rm Im}\,t<\beta$, these singularities appear on the second sheet of $G^>$, thus making contact with the continuation argument of \cite{Fidkowski:2003nf}.  

In summary, the high-frequency expansion of two-point functions is asymptotic in $1/\omega$ and receives nonperturbative corrections associated with complex-time singularities. These are particularly easy to observe when $d$ and $2\Delta_{\cal O}$ are even integers (which includes the important example of stress tensor correlators in $d=4$), but they are always present in a holographic theory.

%is the imaginary part of the above observable evaluated over the red contour $\mathcal{C}$ in Fig. \ref{fig:contour}. Here $t$ denotes a small time scale compared to the scrambling time. The contour $\mathcal{C}$ can be deformed to pick up the poles of the two-point function as in Fig. \ref{fig:contour}.

%For the current-current correlator, the maximum contribution comes from the vacuum contribution as shown in Fig. \ref{fig:current}, while the subleading contribution arises from the two poles at $  t_\pm = \pm 1 + \ic$, which are the two leading geodesics in Fig \ref{fig:geod1}. Our goal in this section would be to evaluate these contributions using the bulk. 

\section{WKB proposal for bouncing geodesics}
\label{sec:proposal}

% We are now ready to state our proposal to account for 
% the contribution of bouncing geodesics to correlators with time separation ${\rm Im} t=\beta/2$. We stress that these geodesics do \emph{not} contribute to the standard two-sided correlator between the left and right Kruskal patches of the black hole geometry, which is given by $G^>$ evaluated in the middle of its analyticity strip $-\beta<{\rm Im} t<0$.  In the time domain the singularities are revealed by analytically continuing outside that strip \cite{}; in our setup this is a natural consequence of the large real frequency.

We are now ready to state our proposal. We will focus on $d=4$, where we can numerically test it, and set $r_h=\frac{4\pi T}{d}=1$ for convenience.

\subsection{Leading reflection coefficient}
\begin{figure}[t!]
    \centering
\includegraphics[width=0.99\linewidth]{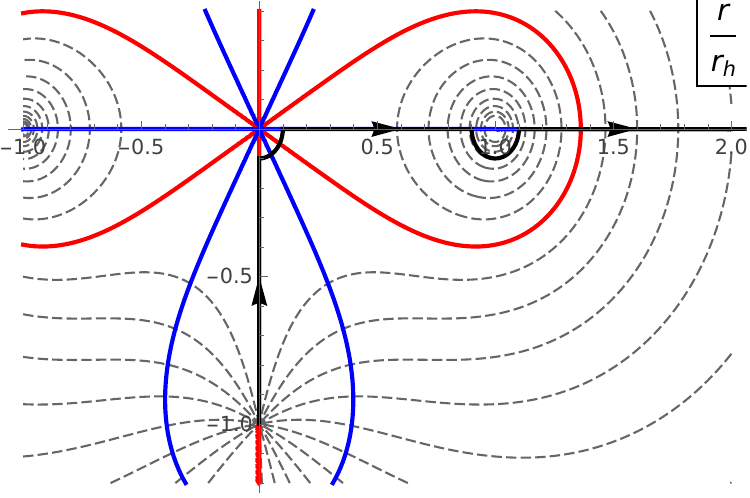}
    \caption{Stoke's lines (in blue, with phases $n\pi/3$ near the origin), along which WKB solutions are oscillatory, and lines of constant WKB phase (solid red and gray dashed). We track $\phi^{\rm steepest}(r)$ from $r=-\ic r_h$ to large $r$ following the solid path with arrows.}
    \label{fig:WKB}
\end{figure}

Working in frequency space, we propose to identify the contribution from the vertical steepest descent contour at the boundary in Fig.~\ref{fig:contour}, which represents a canonical Borel resummation of the leading large-$\omega$ series, with the solution to bulk equations of motion which decays along a steepest-descent contour in $r$. Steepest-descent  lines in the WKB approximation are shown in Fig.~\ref{fig:WKB}. Starting from large real $r$, we observe that the lines follow a large circle which ends at the attractor point $r=-\ic$, with infalling solution $\overleftarrow{\phi_{\omega}}(r)$ initially decaying along this line. Hence, the solution $\phi_\omega^{\rm steepest}(r)$, defined by being regular at $-\ic$, must \emph{almost} coincide with the infalling solution:
\begin{equation} \label{small schematic}
  \overleftarrow{\phi_{\omega}}(r) = \phi_\omega^{\rm steepest}(r) + \mbox{(exponentially decaying in $\omega$)}.
\end{equation}
The question is how to calculate the decaying piece, which, based on Fig.~\ref{fig:contour}, we identify with nonperturbative effects.

%Propagation of a particle from the left or right boundary to near the singularity can be calculated using standard WKB approximations, hence the essence of the question is to identify the correct boundary condition to impose at the singularity, where WKB breaks down.

We must connect $r=-\ic$ to the horizon $r=1$ where the boundary condition on $\overleftarrow{\phi_{\omega}}(r)$ is imposed.  Taking inspiration from \cite{Motl:2003cd}, we follow in turns the imaginary and real axes, as shown in Fig.~\ref{fig:WKB}. There the WKB solution is, respectively, purely increasing and purely oscillatory.
The axes cannot be connected directly: they are separated by a Stokes line along which the WKB solutions exchange dominance. Rather, we proceed through the small-$r$ region. The WKB approximation breaks down when $r^3\omega\sim1$ and the $\phi''$ and $\omega^2$ terms in \eqref{eom} compete.
Introducing $x=r^3\omega/3$, the wave equation in this scaling limit (large $\omega$ with fixed $x$) becomes
\begin{equation} \label{eom x}
  0= \partial_x^2\phi+\frac{1}{x}\partial_x\phi+\phi
    +\mathcal{O}(q^2\omega^{-2/3},m^2\omega^{-4/3}) .
\end{equation}
Recall that we are taking $\omega$ large with $q$ and $m$ fixed. 
Luckily, this  can be solved exactly.
The desired solution, which decays along negative imaginary $r$, can be written on the real axis as a sum of oscillatory Bessel/Hankel functions:
\begin{equation} \label{phi steepest}
    \phi_\omega^{\rm steepest}(r)\propto K_0(-\ic x) =
\frac{\pi}{2\ic} \left(H^{(1)}_0(x) + 2H^{(2)}_0(x)\right),
\end{equation}
where the equality can be verified by series expansion in $x$ and continuing termwise through the $r$ fourth quadrant.
Parametrizing the ratio of infalling and outgoing components near the singularity by a \emph{reflection coefficient} $R$:
\begin{equation}\label{R def}
    \phi_\omega^{\rm steepest}(r) \propto H^{(1)}_0(x) -  R H^{(2)}_0(x) \quad \mbox{(near $r=0$),}
\end{equation}
we conclude from \eqref{phi steepest} that $R(\omega)=-2+\mathcal{O}(\omega^{-2/3})$.
The number ``$-2$'' will be numerically confirmed below.

The final step is to evolve the solution \eqref{R def} along the real axis toward the AdS boundary.
We start from the region $\omega^{-1/3}\ll r\ll 1$ where $x$ is large and each Hankel function can be matched to a WKB solution. We then evolve to large $r$ by integrating the WKB phase, passing below the horizon at $r=1$, until the region $1\ll r\ll \omega$ where we can match with boundary Hankel functions discussed in Appendix~\ref{app:expansions}.
The relevant limits of the WKB phase are
\begin{equation}
 \int_0^r \frac{\d r}{r^2f(r)}    
\to \left\{\begin{array}{ll}
-r^3/3 ,
& r\to 0 ,
\\
\frac{\beta}{4}(1+\ic)-1/r ,\quad & r\to\infty .
\end{array}\right.
\end{equation}
These limits match with the near-singularity and near-boundary Hankel functions, which are thereby connected:
\begin{equation}\begin{aligned} \label{matching chain}
%\phi_0^>\supset
&
H_0^{(1)}(r^3\omega/3) \xleftrightarrow[{\rm small}\,r]{} \frac{\sqrt{6}r^{-\frac32}}{\sqrt{\ic\pi\omega}}
e^{-\ic\omega\int_0^r \frac{\d r}{r^2f(r)}}
\nonumber\\&\quad
\xleftrightarrow[{\rm large}\,r]{}\sqrt{3} e^{\frac{\beta \omega}{4}(1-\ic)+\frac12\ic\pi\nu}
 r^{-2}H^{(1)}_\nu(\omega/r).
%\frac{(-\ic)^\nu}{\sqrt{3}} 
\end{aligned}\end{equation}
The real part of the exponent
coincides with the change in the infalling mode $\phi\propto (r-1)^{\frac{-\ic \beta\omega}{4}}$ around $r=1$ (related to the Schwinger-Keldysh prescription of \cite{Glorioso:2018mmw, Chakrabarty:2019aeu, Bu:2020jfo, Loganayagam:2022zmq}). The continuation of the outgoing solution $H_0^{(2)}$ is similar but with the opposite exponent. Thus \eqref{R def} evolves as $r\to\infty$ to:
\begin{equation} \label{steepest AdS}
    \phi_\omega^{\rm steepest}(r)\!\to
r^{-2}\!\left[H_\nu^{(1)}(\omega/r)-\!R e^{-\frac{\beta \omega}{2}(1-\ic)-\ic\pi\nu} H_\nu^{(2)}(\omega/r)\right]\!,
\end{equation}
where we fixed a convenient overall normalization.
To the accuracy of our calculation, the first term coincides with the true infalling solution. Hence, as anticipated in \eqref{small schematic}, the steepest-descent and infalling solutions differ by an exponentially small correction:
\begin{equation} \label{infalling from R}
    \overleftarrow{\phi_\omega}(r) \to
    \phi_\omega^{\rm steepest}(r) + R e^{-\frac{\beta \omega}{2}(1-\ic)-\ic\pi\nu} r^{-2}H_\nu^{(2)}(\omega/r).
\end{equation}
% up to expected nonperturbative corrections of order $e^{-\beta\omega}$.
According to our proposal, the first contribution represents the canonical Borel resummation of the $1/\omega$ asymptotic expansion.
Hence, applying the holographic recipe \eqref{recipe}, we have obtained the following transseries solution:
\begin{equation}\label{Gret transseries}
  G_{\rm ret}(\omega)= \omega^{2\nu}\!\left[\begin{array}{l} \frac{e^{-\ic\pi\nu}}{-2\sin(\pi\nu)}+\mathcal{O}(\omega^{-4}) \\
+\ic e^{-{\beta \omega \over 2}(1-\ic)-\ic\pi \nu}R(\omega)\times (1+\mathcal{O}(\omega^{-4}))
\\ +\cdots\end{array}\right]
% \end{equation}
\end{equation}
where $\cdots$ represent subleading paths which we expect to be more suppressed, $\propto e^{-\beta\omega}$. This is the main result of this paper. Here we chose the normalization $C$ in \eqref{recipe} so that the zero-temperature spectral density is $2{\rm Im} G_{\rm ret}^{\rm vac}= \omega^{2\nu}$, and we recall that $\nu=\Delta_{\cal O}-\frac{d}{2}$.

Using Fourier transforms from Appendix \ref{app:Fourier},
an equivalent statement of \eqref{Gret transseries} is that, in the time domain, the retarded function contains a specific singularity at $t_*=\frac{\beta}{2}(1+\ic)$ [see \eqref{Gt asympt}]:
\begin{equation} \label{Gret sing}
G_{\rm ret}(t,q{=}0)\to -2\ic \times
\frac{\Gamma(2\nu{+}1)}{2\pi(\ic(t{-}t_*))^{2\nu+1}} +\mbox{(less singular),}
\end{equation}
in the same normalization.
This controls, for example, the large-order behavior of the OPE coefficients $a_n$ [see \eqref{large n app}].

\subsection{Subleading reflection coefficient} \label{app:reflection}

The series \eqref{Gret transseries} with $R(\omega) \approx −2$ is our main result.  Here we explain how perturbation theory near the singularity can be used to obtain the first subleading $\sim \omega^{-4/3}$ correction to $R$ (see \eqref{R NLO}), and also explain the size of the error terms in \eqref{Gret transseries} in terms of contributions from other regions. 
%Here we detail our analysis of the leading perturbative $1/\omega^{4/3}$ correction to the bouncing geodesic contribution outlined above.

The correction will come entirely from near the singularity.
We will focus here on the case of zero spatial momentum, $q=0$. %, although the analysis of the $q^2/\omega^{2/3}$ corrections mentioned in the test would proceed similarly.
In terms of the scaling variable $x=\omega r^3/3$ (we set $r_h=1$ here) the radial equation reads
\begin{equation} \label{DE x app}
     \phi''+\frac{1}{x}\phi'+\phi =
%    -\frac{q^2}{(3x)^{4/2}\omega^{2/3}}
   \frac{12x\phi'-(18x^2+m^2)\phi}{(3x)^{2/3}\omega^{4/3}}
+\mathcal{O}(q,\omega^{-8/3}).
\end{equation}
The left-hand side is the equation studied in the text. Its zeroth-order solution of interest decays for negative imaginary $r$: $\phi^{(0)} = K_0(y)$ from \eqref{phi steepest}, where one initially takes $y=\ic^3x$ real.
The perturbed solution $\phi=\phi^{(0)}+\phi^{(1)}+\cdots$ can be found using Green's function method
%\begin{align}
\begin{equation}\begin{aligned}
     \phi^{(1)}(y') &=
     {-}\!\int_0^\infty\!y\d y G(y',y)
\frac{12y\partial_y\phi^{(0)}{+}(18 y^2{-}m^2)\phi^{(0)}}{(3y)^{2/3}\omega^{4/3}},
\\
G(y',y)&=%\! {-}y'\bigl[
\theta(y{-}y')K_0(y)I_0(y')+(y\leftrightarrow y')
%\theta(y'{-}y)I_0(y)K_0(y')%\bigr]
\end{aligned}\end{equation}
%\end{align}
 where we have selected the Green's function with the same $y\to\infty$ boundary condition as $\phi^{(0)}$.
Although we were not able to evaluate this integral for generic $y$, its $y'\to 0$ limit can be evaluated as a complete Bessel integral:
\begin{align} \label{lim y0 app}
\lim_{y\to 0}\phi^{(1)}(y) &= \log\frac{2}{y}
-\gamma_E + \frac{C}{\omega^{4/3}}\left(m^2+\frac{24}{7}\right) +\mathcal{O}(\omega^{-8/3}),
\nonumber\\
    C&\equiv \frac{1}{3^{2/3}}\int_0^\infty y^{1/3} \d y K_0(y)^2 = \frac{2^{2/3}\pi^2\Gamma(\frac23)}{3^{1/6}\Gamma(\frac{1}{6})^2} .
\end{align}
As done above, we now continue $r$ through the fourth quadrant, which replaces the logarithm by $\log\frac{2}{x}-\frac{3\ic\pi}{2}$, which we can then use as an initial condition for the differential equation \eqref{DE x app} at real $x$.
To evolve these data to large $x$, we follow the same procedure and use Green's functions which have, respectively, the same asymptotics as $H^{(1)}(x)$
and $H^{(2)}(x)$.  The large-$x$ asymptotics are then expressed in terms of complete integrals over oscillatory Bessel functions, but we find that all integrals can again be expressed in terms of the same constant $C$.
Omitting details, we find
\begin{equation}\begin{aligned}
   \!\!\!\lim_{x\to\infty}\phi&\propto
  \!\bigl(1{+} \ic\omega^{-4/3}\delta\varphi\bigr)H^{(1)}_0(x)
\\
&+\bigl(1{-}\ic\omega^{-4/3}\delta\varphi\bigr)H^{(2)}_0(x)\bigl[2+\omega^{-4/3}R^{(1)}e^{\ic\pi/3}\bigr],
\end{aligned}\end{equation}
up to corrections suppressed at large $x$ or by $1/\omega^{8/3}$.
Here $R^{(1)}$ is a constant given below and $\delta\varphi=\frac17(3x)^{7/3}+\frac18(15+4m^2)(3x)^{1/3}$
is a phase which does \emph{not} decay at large $x$.
However, the large-$x$ behavior has to be matched with the WKB ansatz expanded to the equivalent order, which at generic $r$ contains the terms
\begin{equation}
 \phi(r) \!\propto\! r^{-\frac{3}{2}}
 \!\exp\!\left[\ic \omega\!\!  \int { \diff r \over r^2 f(r)}+\ic\frac{(15+4m^2)r^4{-}3}{8\omega r^3} \!+\mathcal{O}(\omega^{-2})\right]\!.
\end{equation}
 % \int \diff r \frac{(d-1) r^2 \left((d+1) f(r)+2 r f'(r)\right)+4 \left(q^2+m^2 r^2\right)}{8 r^2} + \dots }
By expanding this in the scaling region $\omega^{-1/3}\ll r\ll 1$ and expanding the Hankel functions to subleading order we find a precise agreement with $\delta\phi$, enabling a $r$-independent matching to the WKB ansatz, as it should.
This enables an unambiguous definition of the reflection coefficient [generalizing \eqref{R def}] at this order:
\begin{equation}\begin{aligned} \label{R NLO}
    R&={-}2{-}\frac{e^{\ic\pi/3}R^{(1)}}{(\beta\omega/\pi)^{4/3}}
+\mathcal{O}(\frac{1}{\omega^{8/3}}) ,\\
   R^{(1)}&=
\frac{3\sqrt{3} C}{\pi}
\frac{24+7m^2}{7} =\frac{\pi (24+7m^2)\Gamma(\frac23)}{2{\times} 18^{1/3}\Gamma(\frac16)\Gamma(\frac{13}{6})} .
\end{aligned}\end{equation}
where we restored $r_h=\pi/\beta$ and $m^2=\Delta(\Delta-4)$.
This nontrivial prediction is verified numerically below.

In the above we can see that the $1/\omega$ terms present in the WKB exponent for generic $r$ neatly canceled when matching the regions, and we observe similar cancellations near the AdS boundary. Hence we expect that the corrections to the boundary correlator consist of a double series combining  powers of $1/\omega^{4/3}$ from the singularity and powers of $1/\omega^4$ from the boundary, with no other powers of $\omega$.

At nonzero but small $q$, we expect the calculation of $q^2/\omega^{2/3}$ terms to be qualitatively similar.  On the other hand, further orders in $1/\omega$ might be significantly more difficult to calculate since perturbation theory will produce iterated integrals over products of Bessel functions.

\section{Numerical tests}

We can test \eqref{Gret transseries} in two ways. First, for the special values of $\Delta$ for which the leading series for ${\rm Im}\,G_{\rm ret}$ terminates,
we can simply measure $R$ by plotting $G^>-G^>_{\rm vac}$.
Second, for general $\Delta$, we can compare the $t$-plane singularities predicted by \eqref{Gret sing} with the large-order behavior of the OPE coefficients $a_n$ in \eqref{Gret t}. We discuss these in turns.

Before going into the details, we describe the results of the first test where we set $\Delta=4$ in \eqref{Gret transseries} (still in $d=4$). We observe from \eqref{Gret omega} that the leading transseries for the imaginary part terminates and consists of a single term, leading to the large-$\omega$ prediction
\begin{equation}\begin{aligned} \label{large omega ansatz}
  &2{\rm Im} G_{\rm ret}^{\Delta=4}(\omega,q{=}0)\to \omega^4+\omega^4 e^{-\frac{\beta\omega}{2}}\times
 \\ 
 &\times
 {\rm Re}\, e^{\ic \frac{\beta\omega}{2}}
\left[ c_0 + %\cos(\tfrac{\beta\omega}{2})+
\frac{c_1 e^{\ic\phi_1}}{(\beta\omega/\pi)^{4/3}}+
\frac{c_2 e^{\ic\phi_2}}{(\beta\omega/\pi)^{8/3}}+\ldots\right]
\end{aligned}\end{equation}
with $c_0=2R|_{\omega=\infty}=-4$ and known values for $c_1$ and $\phi_1$ (see Table~\ref{tab:fit1}).  Note that the equation of motion for a scalar with $\Delta=4$ is the same as that of the shear mode $T^{xy}$ of the stress tensor studied in \cite{Teaney:2006nc}, who observed exponential decay in $\omega$ after subtracting the vacuum. %Hence, equivalently, \eqref{prediction T} gives a prediction for the stress-tensor correlator $2{\rm Im} G_{\rm ret}^{xy,xy}$.
Figure~\ref{fig:stress tensor} confirms our description of this decaying term since the remainders after subtracting the $c_0$ and $c_1$ terms become up to 5 orders of magnitude smaller. Numerical fits to the coefficients $c_0$ and $c_1$, shown in Table~\ref{tab:fit1}, also demonstrate precise quantitative agreement.

The second check is to compare the $t$-plane singularity in \eqref{Gret sing} with the large-order behavior of the coefficients $a_n$ in \eqref{Gret t}.  These coefficients can be calculated efficiently (at least for even $d$) for any $\Delta$ using the algorithm in Appendix~\ref{app:expansions}, which is a momentum space version of the algorithm of \cite{Fitzpatrick:2019zqz}.
By Fourier transforming \eqref{Gret transseries} one obtains the detailed prediction in \eqref{large n app}, which again agrees exquisitely with numerics as displayed in Table~\ref{tab:fit2}.

All this confirms the quantitative connections between nonperturbative effects in retarded functions at large real frequencies, complex-time singularities in $G_{\rm ret}(t)$, and the  reflection coefficient [see \eqref{R def} and \eqref{R NLO}] controlling the contribution of null geodesics bouncing off the singularity.

\subsection{Example data for $\Delta=4$}

Here we detail numerical checks performed
on the scalar correlator with $\Delta=4$ in $d=4$, which as just mentioned is equivalent to the so-called shear channel stress tensor correlator $T^{xy}$.

It is relatively straightforward to evaluate the frequency space correlator by numerically integrating the radial equation.
Working in a variable $u=1/r^2$, we compute series expansions around $u=0$ and $u=1$ and match them at the midpoint where each series converges exponentially fast.

\begin{figure}[t!]
    \centering
\includegraphics[width=1.0\linewidth]{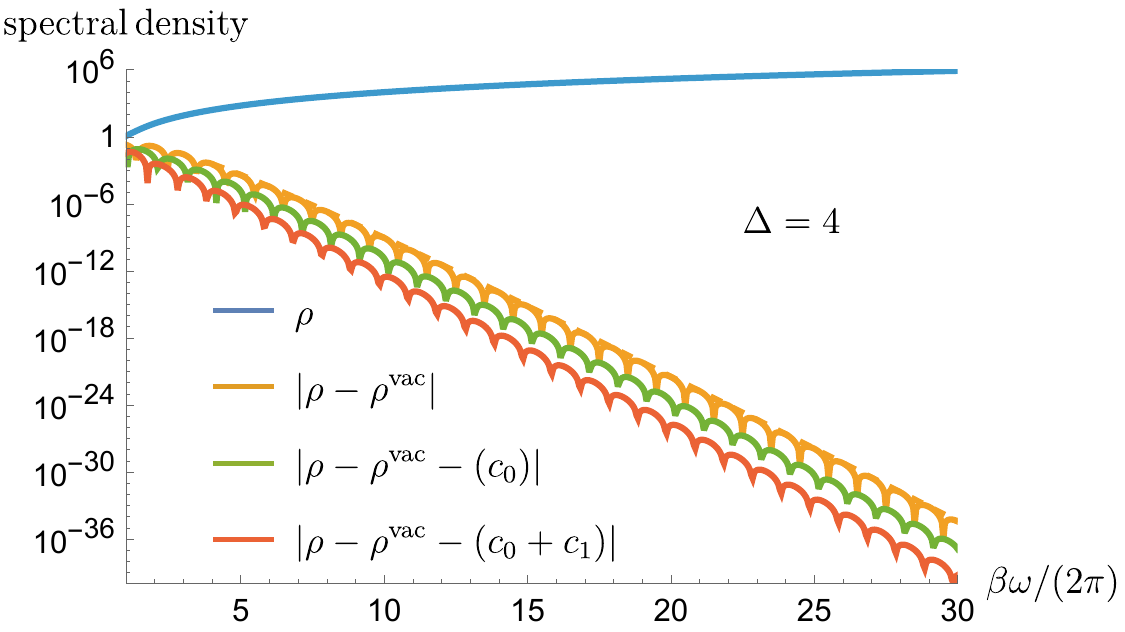}
    \caption{The spectral density $\rho=2{\rm Im} G_{\rm ret}(\omega,q=0)$ for a scalar with $\Delta=4$, showing the decreasing residuals after subtracting various numbers of terms in \eqref{large omega ansatz}. A dashed line atop the second (yellow) curve displays the trend $4\omega^4e^{-\frac{\beta\omega}{2}}$.}
    \label{fig:stress tensor}
\end{figure}

In Fig.~\ref{fig:stress tensor} we display this numerical result, along with the residual after subtracting the vacuum term ($\omega^4$) and either the $c_0$ or both $c_0$ and $c_1$ terms in \eqref{large omega ansatz}. Each subtraction is seen to further decrease the residual, as it should, with a net reduction of up to 5 orders of magnitude at $\frac{\beta\omega}{2\pi}=30$, precisely confirming their values.

We also fitted the correlator to the ansatz \eqref{large omega ansatz}, observing extremely precise agreement with the predictions for $c_0$ and $c_1$ from \eqref{Gret transseries} as shown in Table.~\ref{tab:fit1}.
To perform this fit we kept additional subleading coefficients up to $c_4$ in the ansatz, not shown here. Note that agreement for the phases is somewhat trivial since they are fully determined by the $\omega$ exponents together with the reality of $G_{\rm ret}(t,q{=}0)$ at real times; the fitted value of $c_2$ may be interesting for future work.

\begin{table}[t!]
    \centering
    \begin{tabular}{|c|c|c|}
    \hline Coefficient & Prediction & Fit \\\hline
$c_0$ & $-4$ & $-3.9999999999994$\\
$c_1$ & $2R^{(1)}\big|_{m=0}\approx 6.46639430\ldots$ & 
$6.46639431$\\
$\phi_1/\pi$ & $-2/3$ & $-0.666666668$\\
$c_2$ & unknown & $6.93972$\\
$\phi_2/\pi$ & $-1/3$ & $-0.3333331$\\\hline
    \end{tabular}
    \caption{Coefficients in the large-$\omega$ expansion of the $\Delta=4$ spectral density in \eqref{large omega ansatz}. The fit uncertainties are estimated to be on the last digits, based on comparing the fit using $\frac{\beta\omega}{2\pi}\in [20,30]$ versus $[20,25]$.}
    \label{tab:fit1}
\end{table}

The coefficient $c_0=-4$ in Table~\ref{tab:fit1} has a simple physical interpretation: the leading nonperturbative correction to $2{\rm Im}G_{\rm ret}$ at large frequencies is the sum of two reflected geodesics (bouncing off, respectively, from the future or past singularity), each weighted by a reflection coefficient $R=-2$. This value helped us single out and identify the derivation presented in Sec.~\ref{sec:proposal}.

Another test is to compare the large-order behavior of the single-trace OPE coefficients determined from Appendix~\ref{app:expansions} to the prediction in \eqref{large n app}. Again we focus on spatial momentum $q=0$ ($\gamma=1$).
The advantage of this test is that it can be carried out for any $\Delta$.
Inspired by the structure of \eqref{large n app}, we fit the $a_n$'s obtained from the differential equation to a double expansion in $1/n$ and $1/n^{4/3}$:
\begin{equation} \label{large n ansatz}
    \frac{2\pi a_n}{(-4)^n n^4} \to d_0 + \frac{d_1}{n} + \frac{d_{4/3}}{n^{4/3}}+\frac{d_{2}}{n^2}+
    \frac{d_{7/3}}{n^{7/3}}+\frac{d_{8/3}}{n^{8/3}} +\ldots
\end{equation}
According to \eqref{large n app}, we expect the coefficients
of integer powers of $1/n^{4/3}$ to determine all other coefficients through $r_{p,\nu}$ factors (whose purpose is to cancel $1/\omega$ correction in frequency space), but in our fit we treat them as independent.  

For $\Delta=4$ we obtained 500 $a_n$ terms and used $n\in [200,500]$ to fit with \eqref{large n ansatz} including terms up to $1/n^7$.  The results for the first few coefficients are presented in Table~\ref{tab:fit2}, again revealing precise agreement.

\begin{table}[t!]
    \centering
    \begin{tabular}{|c|c|c|}
    \hline Coefficient & Prediction & Fit \\\hline
$d_0$ & $2048$ & $2047.999999999999999998$\\
$d_1$ & $-5120$ & $-5119.9999999999995$\\
$d_{4/3}$ & $1511.3509453005\ldots$ & $1511.350945300$\\
$d_2$ & 4480 & $4480.000000$ \\
$d_{7/3}/d_{4/3}$ & $-19/9$ & $-2.111111112$ \\
$d_{8/3}$ & unknown & $740.420$ \\
$d_{3}$ & $-1600$ & $-1600.001$\\\hline
    \end{tabular}
    \caption{Coefficients in the large-$n$ behavior \eqref{large n ansatz} of the single-trace coefficients for $\Delta=4$ using $a_n$ with $n\in [200,500]$. The analytic prediction for $d_{4/3}$ is $16(4\pi)^{4/3}R^{(1)}$. Numerical results were truncated to the first figure which changed when modifying the range to $[200,400]$ or changing the ansatz degree.}
    \label{tab:fit2}
\end{table}

This large-order analysis was important in our study for several reasons. First, the value of $d_0$ is equivalent to the leading term $c_0=-4$ discussed above, thus independently confirming that $R\approx -2$ to leading order.  The large-order numerics further indicated that this value was independent of $\Delta$, an important hint of its universal origin from near the singularity.

Second, our initial large-$n$ fits included only powers of $1/n$ (and eventually, logarithms) but this led to poor fits and the extracted values were not stable beyond the leading figures.  Modeling the corrections by powers of $1/n^{4/3}$ immediately stabilized and improved the fits, which suggested the calculation in Sec.~\ref{app:reflection}.

Finally, we also repeated the large-$n$ analysis for many ${\cal O}(1)$ values of $\Delta$, again finding perfect agreement within errors, thus  confirming the simple functional dependence of $R^{(1)}$ on $\Delta$ in \eqref{R NLO}.

\section{Detailed all-order analysis for  R-current correlators}
\label{app:current}

To illustrate the ideas in a case where analytic solution is possible, we look at the radial equation corresponding to the transverse current correlators $G^{xx}$ \cite{Myers:2007we}. We find a different near-singularity scaling limit controlled by Airy instead of Bessel functions, and a corresponding reflection coefficient  $Re^{-\ic\pi\nu}=1$ in agreement with \eqref{chi}.  Here we will also analyze sub-subleading saddles, e.g. multiply-bouncing geodesics.

\begin{figure}[t!]
    \centering
\includegraphics[width=1.0\linewidth]{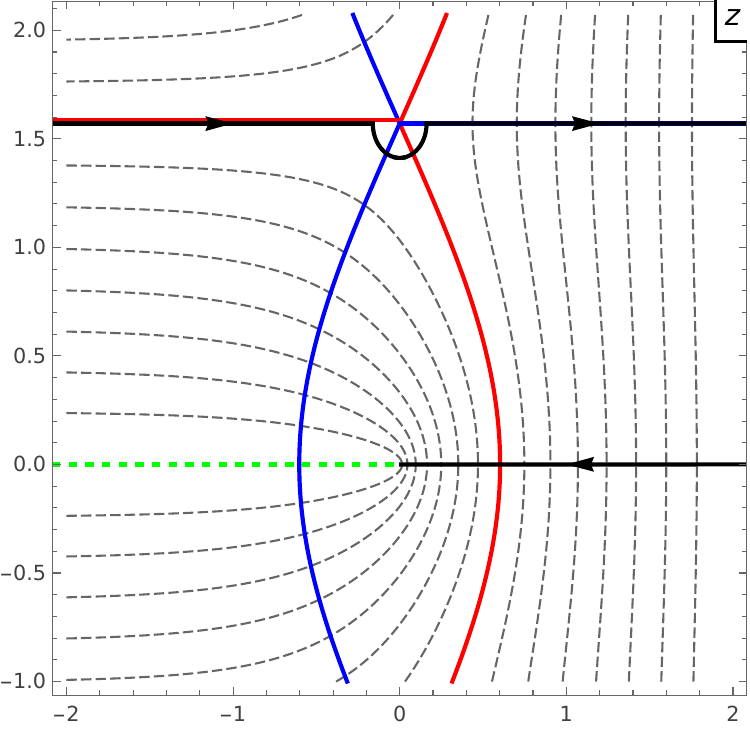}
    \caption{Stokes lines (blue, at angles 0 and $\pm2\pi/3$) where saddles exchange dominance, and    
    anti-Stokes lines (red), where ${\rm Re}\,S=0$, in the complex $z$-plane.
    The AdS boundary is at $z\,{=}\,0$, the black hole singularity at $z\,{=}\,\ic \pi/2$, and the horizon at $z\to{+}\infty$.
    We placed branch cuts of the solution going to the left (in dashed green).
    The black line depicts the contour described in the text.}
    \label{fig:z-plane}
\end{figure}

We analyze the two point functions of transverse $R$-currents $J^x$ with $q=0$ in the planar AdS${}_5$ black hole.
The analysis will be simplified by  defining a ``periodic coordinate'' $z$ by
\begin{equation}
    \tanh(z)=\frac{r_h^2}{r^2}\,.
\end{equation}
The wave equation for the $R$-charge currents \cite{Teaney:2006nc} at $q=0$ in these coordinates becomes
(we set $\beta=2\pi$ in this section)
\begin{equation} \label{chi eom}
    \chi ''(z)+\omega ^2  \coth (z)\chi (z)=0
\end{equation}
where $ \phi(r) =  \frac{\sqrt{r^4 - r_h^4}}{\sqrt{r}}  \chi\left(z\right)$. The complex $r$-plane and $z$-planes are shown in Fig.~\ref{fig:z-plane}. The AdS boundary at $r=\infty$ is mapped to $z=0$ and the horizon at $r=r_h$ is mapped to $z=+\infty$; the singularity at $r=0$ is mapped to $z=\ic \frac{\pi}{2}$ and its images at $z=\ic \frac{\pi}{2} +\ic n\pi$ for $n\in \mathbb{Z}$. In these coordinates one finds an infinite series of simple turning points in the $z$-plane, hence the name. The exact solution to the wave equation with ingoing boundary conditions at the horizon is given by
\begin{align}\label{exact_z_soln}
    &\chi(z)= \frac{\Gamma\!\left(1+\frac{1-\ic}{2}\omega\right)\Gamma\!\left(1-\frac{1+\ic}{2}\omega\right)}
{2^{-\frac{1+\ic}{2}\omega }\Gamma(1-\ic \omega)}
\times\frac{(\coth (z){-}1)^{-\frac{\ic \omega}{2}}}{(\coth (z){+}1)^{\frac{\omega }{2}}}  \nonumber\\
    &\quad\times{}_2F_1\left[1-\frac{1+\ic}{2}\omega,-\frac{1+\ic}{2} \omega ,1-\ic \omega ,\frac{1-\coth (z)}{2} \right],
\end{align}
which we normalized to 1 at the AdS boundary $z=0$. Expanding in this limit we find
\begin{align}
    \chi(z)&=1+z\omega ^2(1- \log (z))+z\kappa_2+O(z^2),\label{frobin}
\end{align}
where $\kappa_2$ gives the retarded Green's function as \cite{Myers:2007we}
\begin{equation}
\begin{aligned} \label{Gxx exact}
    G^{xx}_{\rm ret}(\omega,q{=}0)&={-}\omega^2  \Biggl[
\psi ^{(0)}\left(1+\frac{1-\ic}{2}\omega\right)
+\psi ^{(0)}\left(-\frac{1+\ic}{2}\omega \right) 
\\ &\qquad\quad
+2\gamma_E+\log 2-\omega^{-1}\Biggr]\,.
\end{aligned}\end{equation}
Taking the imaginary part reproduces the spectral density quoted in \eqref{chi}:
\begin{align}
    {\rm Im}\,G_{\rm ret}^{xx}(\omega,q{=}0)=\frac{\pi  \omega ^2 \sinh (\pi  \omega )}{\cosh (\pi  \omega )-\cos (\pi  \omega )}.\label{exact_rho}
\end{align}
In the rest of this section, we will rederive these expressions using WKB methods.

\subsection{Leading WKB approximation}
The leading WKB approximation to differential equations of the form 
\begin{equation}
    \chi ''(z)+Q(z)\chi (z)=0\label{z-wave-eq}
\end{equation}
is given by 
\begin{equation} \label{WKB chi general}
 \!   \chi(z)=\frac{1}{Q(z)^{1/4}}\left(A e^{\ic \int \sqrt{Q(z)} \diff z}+B e^{-\ic \int \sqrt{Q(z)} \diff z}\right).
\end{equation}
In our case, we have $Q(z)=\omega^2 \coth(z)$, and to compute the retarded Green's function, we need to impose ingoing boundary conditions at the horizon $z=\infty$, which sets $B=0$. The WKB solution near the horizon is then given by 
\begin{equation} \label{WKB chi}
    \chi(z)=\frac{e^{\ic S_0}}{(\omega^2\coth[z])^{1/4}}
\end{equation}
where
$S_0=\omega  z-\int_z^{\infty } \omega  \left(\sqrt{\coth (z')}-1\right) \diff z'$ and we have expressed the action in this way to reabsorb the divergent piece into the normalization constant. We wish to know the solution near the boundary at $z=0$. For small $z$, the general solution \eqref{frobin} is readily obtained by a Frobenius expansion.
To connect the WKB solution to the boundary, we need to consider an intermediate matching regime $\frac{1}{\omega^2}\ll z\ll 1$, where the differential equation simplifies to the Bessel form
\begin{equation}
    \chi ''(z)+\frac{\omega ^2}{z} \chi (z)\approx 0.\label{wave_eq_z}
\end{equation}
% and the solutions are given in terms of Bessel functions
% \begin{equation}
%     \chi(z)=a_1 \omega  \sqrt{z} J_1\left(2 \sqrt{z} \omega \right)+2 \ic a_2 \omega  \sqrt{z} Y_1\left(2 \sqrt{z} \omega \right).
% \end{equation}
% Assuming that $\omega$ is real and sufficiently large, we can match the large z asymptotics of the Bessel functions to the WKB solution, and we find
% \begin{align}
%     a_1&=\frac{(1+\ic) \sqrt{\frac{\pi }{2}} e^{\frac{1}{4} \ic \omega  (\pi +\log (4))}}{\sqrt{\omega }},\\
%     a_2&=-\frac{\left(\frac{1}{2}+\frac{\ic}{2}\right) \sqrt{\frac{\pi }{2}} e^{\frac{1}{4} \ic \omega  (\pi +\log (4))}}{\sqrt{\omega }}
% \end{align}
The oscillatory solution in \eqref{WKB chi} then matches specifically to the Hankel function
\begin{align} \label{retarded chi app}
    \chi(z)&\to\sqrt{\pi z}e^{-\frac{1}{4} \ic \omega  (\pi +\log (4))+\frac{3\ic\pi}{4}} H_1^{(1)}\left(2 \sqrt{z} \omega \right).
\end{align}
Expanding near the AdS boundary $z\to 0$ as above, we find the leading approximating to the retarded Green's function
\begin{equation}\begin{aligned}
    G_{\rm ret}^{xx}(\omega,q{=}0)&\approx \omega ^2 (-2 \log (\omega )-2 \gamma_E +\ic \pi ),
\end{aligned}
\end{equation}
which agrees, of course, with the leading term of \eqref{Gxx exact} as $\omega\to\infty$.  The corrections are given by a nontrivial asymptotic series in $1/\omega^4$ with real coefficients. We now show how the method from Sec.~\ref{sec:proposal} captures the nonperturbative corrections to it.

\subsection{First nonperturbative correction}
\label{sec:zpath}

The red path used in Figure~\ref{fig:WKB} is shown in the $z$-plane in Figure~\ref{fig:z-plane}.
The coordinate map $\tanh z = r_h^2/r^2$ sends the point $r=-\ic r_h$ (the attractor point of steepest-descent contours in the complex $r$-plane) to
\begin{equation}
z_* = -\infty.
\end{equation}
As in the text, we denote as $\chi^{\rm steepest}(z)$
the solution which is exponentially decaying as ${\rm Re}\,z_*\to-\infty$; it is approximately equal to the infalling solution at large $\omega$, but is not exactly the same.

To evolve $\chi^{\rm steepest}(z)$ to positive $z$ where we can most easily compare with \eqref{exact_z_soln}, we stay on the line ${\rm Im}\,z=\pi/2$ so as to avoid crossing a Stokes line.
The price to pay is that we hit
%Starting from $z=z_*$, we move along the horizontal line at $\Im z = \pi/2$ toward increasing $\Re z$.  This path represents the analytic continuation of the infalling wave beyond the horizon and parallels the vertical contour used in the Borel analysis of the time integral.
the turning point of the potential $\omega^2\coth(z)$ 
at \begin{equation}
z_s = \ic \frac{\pi}{2},
\end{equation}
which corresponds physically to the curvature singularity $r=0$ in the original radial coordinate.
Near this point the WKB expansion breaks down because $\coth z$ vanishes linearly.  To resolve this region we introduce the scaling variable
\begin{equation}
x = \omega^{2/3}(z-z_s).
\end{equation}
Keeping $x$ fixed as $\omega\to\infty$ transforms the wave equation \eqref{z-wave-eq} into the universal Airy form
\begin{equation}
\chi''(x)+x\chi(x)\approx 0.
\end{equation}
The solution that decays toward negative $x$ (and thus negative $z$) is given by
\begin{equation}\label{airyexp}
\chi^{\rm steepest}(x)\propto \mathrm{Ai}(-x).
\end{equation}
The Airy function is entire, but, as is well known, its asymptotics at positive $x$ are not a simple continuation of the decaying exponential. Rather, it is a combination of two oscillatory exponentials:
\begin{equation} \label{Airy R}
 \mathrm{Ai}(-x)\to 
\frac{e^{\ic \pi/4}}{2\sqrt{\pi}x^{1/4}}
\left(e^{\ic\frac{2}{3}x^{3/2}}-\ic  R e^{-\ic\frac{2}{3}x^{3/2}}\right), \quad R= -1\,,
\end{equation}
where we defined a reflection coefficient $R$ using the same phase convention as in \eqref{R def}.  This solution can be matched to the WKB form \eqref{WKB chi general} in an overlap region where both the WKB and Airy approximations are valid,
\begin{equation}
\omega^{-\frac{2}{3}}\ll \bigl|z-\tfrac{\ic\pi}{2}\bigr|\ll 1,
\qquad
x\gg 1\,.
%\xi \equiv \omega^{2/3}\!\left(z-\tfrac{\ic\pi}{2}\right)\ \ \text{fixed}.
\end{equation}
Continuing each WKB solution and matching to the Bessel functions at $z\to 0$ (the AdS boundary) as in \eqref{retarded chi app} then gives, up to a proportionality constant:
\begin{equation}
    \chi^{\rm steepest}(z) \to
\sqrt{\pi z}\left[
H_1^{(1)}(2\sqrt{z}\omega)+Re^{(-1+\ic)\pi\omega} H_1^{(2)}(2\sqrt{z}\omega)\right]
\end{equation}
which is equivalent to \eqref{steepest AdS} in with $\nu=1$.
Taking $z\,{\to}\, 0$ as in \eqref{frobin} we find
the analog of the transseries
\eqref{Gret transseries} for the retarded function, with $e^{-\ic\pi \nu}R=+1$:
\begin{equation} \label{Gxx transseries}
  G_{\rm ret}^{xx}(\omega,q{=}0)\propto
    \omega^2
\left[\begin{array}{l}
\frac{\ic}{2}-\frac{\log\omega}{\pi} +\mathcal{O}(\omega^{-4}) \\
+\ic e^{-{\beta \omega \over 2}(1-\ic)}\times (1+\mathcal{O}(\omega^{-4}))
\\ +\cdots\end{array}\right].
\end{equation}
In particular, the imaginary part is proportional
to $\omega^2(1+2e^{-\beta\omega/2}\cos(\frac{\beta\omega}{2})+\cdots)$, in precise agreement with \eqref{chi} and \eqref{exact_rho}.
The only difference between the leading nonperturbative contribution to the current and scalar correlators is the value of $R$ [compare \eqref{Airy R} with \eqref{phi steepest}], which is caused by the different behavior near the singularity.

\subsection{All-order nonperturbative corrections}

For the retarded function of currents,
the strategy just described can be implemented to go beyond the leading correction in \eqref{Gxx transseries}. The exact solution to \eqref{chi eom} that is regular at $z\to-\infty$ is
\begin{align}\label{exact_z_steepest}
    &\chi^{\rm steepest}(z)= \frac{\Gamma\!\left(1+\frac{1-\ic}{2}\omega\right)\Gamma\!\left(1+\frac{1+\ic}{2}\omega\right)}
{2^{\frac{1-\ic}{2}\omega }\Gamma(1+\omega)}
\,\frac{(-\coth (z){-}1)^{\frac{\omega}{2}}}{(-\coth (z){+}1)^{\frac{\ic \omega }{2}}}  \nonumber\\
    &\quad\times{}_2F_1\left[1+\frac{1-\ic}{2}\omega,\frac{1-\ic}{2} \omega ,1+ \omega ,\frac{\coth (z)+1}{2} \right].
\end{align}
Upon analytically continuing to positive $z$ with a small positive imaginary part, it becomes a combination of infalling and outgoing modes at the horizon, with the latter suppressed by a relative $e^{-\beta\omega/2}$ as anticipated in \eqref{small schematic}. Let us quantify the remainder more precisely.

A good way to impose the infalling condition at the horizon ($z\to +\infty$) is to impose the correct monodromy $\sim e^{-\beta\omega/2}$ under $z\mapsto z+\ic\pi$ at positive $z$.  Since the similar shift on the left only rescales $\chi^{\rm steepest}$ by a phase $e^{\ic\beta\omega/2}$, this can be phrased in terms of two analytic continuations of the same function:
\begin{equation}
 \chi(z)\propto \chi^{\rm steepest,\rotatebox[origin=c]{0}{$\curvearrowright$}}(z)+
 e^{-\frac{1+\ic}{2}\beta\omega}\chi^{{\rm steepest},\rotatebox[origin=c]{180}{$\curvearrowleft$}}(z)
\end{equation}
where $\chi^{{\rm steepest},\rotatebox[origin=c]{180}{$\curvearrowleft$}}$ is defined by continuing slightly above the image of the AdS boundary at $z=-\ic\pi$, to ensure the correct periodicity on the right. However, because the solution has no singularities between 0 and $-\ic \pi$, this is the same as continuing slightly below the origin. The difference between the two continuations is thus regular at the AdS boundary $z=0$:
\begin{align} \label{chi reg}
\chi^{\rm reg}(z) &\equiv  \chi^{\rm steepest,\rotatebox[origin=c]{0}{$\curvearrowright$}}(z)
-\chi^{\rm steepest,\rotatebox[origin=c]{180}{$\curvearrowleft$}}(z)
\nonumber\\
&= \ic e^{\ic z\omega}\pi\omega^2(1-e^{-2z})
\nonumber\\&\hspace{-12mm}\times
{}_2F_1\left[1+\left(\frac{1-\ic}{2}\right) \omega ,1-\left(\frac{1+\ic}{2}\right) \omega,2,1-e^{-2z}\right]\!.
\end{align}
Combining the previous two equations we can write the infalling solution exactly as
\begin{equation} \label{chi from steepest app}
    \chi(z) = \chi^{\rm steepest,\rotatebox[origin=c]{0}{$\curvearrowright$}}(z)
+ \frac{1}{e^{(1-\ic)\beta\omega/2}-1}\chi^{\rm reg}(z)\,.
\end{equation}
The second term is exponentially small at large real $\omega$, confirming \eqref{small schematic}.
The above is normalized to $1+\mathcal{O}(z)$ at the AdS boundary, and following \eqref{frobin} we get the exact retarded function in the form
\begin{equation}
\begin{aligned} \label{Gxx exact 2}
    G^{xx}_{\rm ret}(\omega,q{=}0)&\!=\!{-}\omega^2  \Biggl[
\psi ^{(0)}\!\left(\!1+\frac{1-\ic}{2}\omega\!\right)
+\psi ^{(0)}\!\left(\!1+\frac{1+\ic}{2}\omega\! \right) 
\\ &
+2\gamma_E+\log 2-\omega^{-1}-\ic\pi\Biggr]
+ \frac{2\pi \ic\,\omega^2}{e^{(1-\ic)\beta\omega/2}-1}\,.
\end{aligned}\end{equation}
This can be verified to be equivalent to \eqref{Gxx exact} using a polygamma identity. The upshot of this form is that
the square bracket, which originates from the first term of \eqref{chi from steepest app}, can be interpreted as a canonical Borel resummation of the large-$\omega$ asymptotic series for real $\omega$; this interpretation is further confirmed in Appendix \ref{Borel R}.  The second term captures the genuine
nonperturbative corrections. We see that they naturally take the form of a geometric series, representing multiple reflections off the singularity and AdS boundary.

For general correlation functions, the step leading to \eqref{chi reg} would be significantly more complicated
since the solutions will not be analytic near the black hole singularity at $z=\frac{-\ic\pi}{2}+\ic \pi n$.
The trivial monodromy there is the main simplifying feature of the $G^{xx}(\omega,q{=}0)$ correlator.

\section{Discussion}

\begin{figure}[t]
    \centering
\begin{tikzpicture}[scale=0.5]
\draw  (-1,0.65)--(-1,0) -- (5,0) -- (5,-0.3) -- (-1,-0.3) -- (-1,-1.6) -- (5,-1.6) -- (5,-1.9) -- (-1,-1.9) -- (-1,-2.55);
\draw  (-1.1,0.65)--(-0.9,0.65) (-1.1,0.72)--(-0.9,0.72);
\draw  (-1.1,-2.55)--(-0.9,-2.55) (-1.1,-2.62)--(-0.9,-2.62);
 \node [left] at (-1,-1.0) {\small $\frac{\ic\beta}{2}$}; \node [left] at (-1,-2.3) {\small $\frac{\ic\beta}{4}$}; \node [left] at (-1,0.3) {\small $\frac{\ic\beta}{4}$};
\filldraw (3.0,0) circle (2pt); \filldraw (0.5,-0) circle (2pt); 
\draw (6,0.35) -- (6,-0.15) -- (6.5,-0.15);
\node [above] at (6.25,-0.15) {$t$}; 
\node [right] at (5.25,-1.25) {$\quad \mapsto$}; 
\end{tikzpicture}
\begin{tikzpicture}[scale=0.5]
\draw  (-1,0.65)--(-1,0) --(3.45,0) (3.75,0)-- (5,0) -- (5,-0.3) -- (-1,-0.3) -- (-1,-1.6) -- (5,-1.6) -- (5,-1.9) -- (-1,-1.9) -- (-1,-2.55);
\draw (3.45,0) -- (3.45,0.65) (3.75,0) -- (3.75,0.65) (3.45,-2.3) -- (3.45,-2.55) (3.75,-2.1) -- (3.75,-2.55) (2.5,-2.1) -- (3.75,-2.1) (2.5,-2.3) -- (3.45,-2.3) (2.5,-2.1) --(2.5,-2.3);
\draw  (-1.1,0.65)--(-0.9,0.65) (-1.1,0.72)--(-0.9,0.72);
\draw  (-1.1,-2.55)--(-0.9,-2.55) (-1.1,-2.62)--(-0.9,-2.62);
 \node [left] at (-1,-1.0) {\small $\frac{\ic\beta}{2}$}; \node [left] at (-1,-2.3) {\small $\frac{\ic\beta}{4}$}; \node [left] at (-1,0.3) {\small $\frac{\ic\beta}{4}$};
\filldraw (3.0,-2.1) circle (2pt); \filldraw (0.5,0) circle (2pt); 
\end{tikzpicture}
     \caption{Analytic continuation of the single-sided correlator in the thermofield double state to the configuration which displays the bouncing geodesics. The two initial timefolds represent the left and right exteriors of the Kruskal geometry.} \label{fig:crrr}
     %sided thermal correlator to a double-sided correlator which captures geodesics reflected off the singularity.} \label{fig:crrr}
\end{figure} 

As in the Introduction, we emphasize that our analysis describes measurements done in the \emph{exterior} of a black hole: the ``interior'' gets generated by analytic continuation of the exterior geometry along a steepest-contour path. All effects of the singularity are contained in the reflection coefficient $R(\omega)$ in \eqref{Gret transseries} \footnote{
The reflection coefficient is expected to receive stringy corrections at frequencies higher than those we considered; see, for example, \cite{Zigdon:2024ljl}.}. As was also stressed in the Introduction, there is no obvious connection between these data and the experience of an infalling observer.

%this data is not necessarily relevant to the experience of an infalling observer.

We elaborate on this point in Fig.~\ref{fig:crrr} by showing the analytic continuation to ${\rm Im}\,t=+\ic\beta/2$, where the signal is found, of the path integral which calculates the correlation function.  The analytic continuation from the right exterior of the Kruskal geometry appears to spawn a copy of the left exterior that is distinct from the ``original'' left region.  For example, modifying the thermofield double state by inserting a unitary to add infalling matter in the original left region would have strictly no effect on its copy. It would also not affect the spectral density measured in the right exterior, as is also clear from the commutativity of left and right operators.

The continued contour in Fig.~\ref{fig:crrr} explicitly violates the Kontsevich-Segal criterion \cite{Kontsevich:2021dmb, Witten:2021nzp} for convergence of the path integral: it features backward Euclidean time evolution. This simply means that there is no simple axiomatic reason for the analytic continuation of $G^>(t)$ to exist. It is unclear whether this complex saddle point can be precisely defined at finite $N$.  Finite-$N$ effects, which are beyond the accuracy of our calculations, could thus modify the exponential decay exemplified in Figs.~\ref{fig:current} and \ref{fig:stress tensor} at sufficiently large frequencies,
%.  A concrete question is whether they suppress or enhance
either suppressing or enhancing the (already exponentially small) signal from the bouncing geodesics.

%At frequencies larger than considered in this work, the typical scale $r\sim \omega^{-1/3}$ probed by the solution will reach the bulk string scale; we expect these corrections to 
% measurement where the interior geometry arises through an analytic continuation of an exterior measurement. The large-frequency Fourier transform makes the exterior problem mathematically equivalent to a geodesic problem in the interior of the eternal black hole, but it does not imply that the metric is directly probed by an infalling observer.

% Our results identify contributions from geodesics reflected off the singularity as the vacuum-subtracted spectral density. The universality of the dips observed in the spectral density supports this interpretation: while higher-derivative and stringy corrections affect the reflection coefficients of these geodesics but leave the qualitative structure of the spectral density intact.

From the field theory perspective, the \emph{absence} of other nonanalyticity in the strip $|{\rm Im}\,t|<\beta/2$ is notable. By conformally mapping the strip to the unit disk via $z=\tanh(\pi t/(2\beta))$,
 %$t=(2\beta/\pi)\tanh^{-1}(z)$,
convergence of the resulting series for $|z|<1$ implies that the small-$t$ OPE series of the retarded function can be resummed to reach late times. Fourier-transforming the KMS condition \eqref{kms0}, this would also explicitly reconstruct all Wightman functions from the retarded function's OPE; this could be relevant for  bootstrap approaches \cite{Buric:2025fye, Buric:2025anb,Barrat:2025nvu}.
Given the difficulty of producing complex-time singularities at weak coupling, it is tempting to conjecture convergence of the OPE for $|z|<1$ in any relativistic thermal field theory.
%, which we have seen is saturated by single-traces at large-$N$,  It would be interesting to contrast this approach with the double-trace bootstrap recently discussed in \cite{Buric:2025fye, Buric:2025anb,Barrat:2025nvu}. 

Our analysis has been limited to zero spatial momentum.
Assuming that the Fourier transform to position space is controlled 
by the $q$-dependence of the WKB phase, $\omega\Delta t\supset |q|^{\frac32}/\omega^{\frac12}$, one would expect $|q|\sim \omega^{1/3}$ to dominate.  Unfortunately, this is exactly where the corrections to \eqref{eom x} become important, invalidating the use of $R\approx -2$ within the Fourier transform.
In general $d$, there is a nontrivial large-$\omega$ scaling regime where $q\sim \omega^{\frac{d-2}{2(d-1)}}$ and $m\sim\omega^{\frac{d}{2(d-1)}}$ and all effects compete near the singularity: 
\begin{equation}
 r^2\phi'' + r\phi' +(\omega^2r^d + m^2 r^2+q^2)r^{d-2}\phi\approx0.
\end{equation}
Calculating the reflection coefficient for this equation would enable Fourier transforming our results to position space
for general $d$ and mass.

Several open questions remain. Can the reflection coefficient at the singularity be related to a final-state projection in black hole quantum mechanics?
Since the contour in Fig.~\ref{fig:WKB} runs along the imaginary $r$ axis, where the radial coordinate plays the role of a time coordinate, does that mean we can interpret quantities in the dual CFT in terms of the  cosmology that lives behind the event horizon? 
How will our analysis change when extended to black holes in asymptotically flat spacetime (see also \cite{Basha:2018bvi})?
Do higher-point correlators allow us to gain further insight into the black hole interior (see also \cite{Chakravarty:2025ncy})?
For black holes with both inner and outer horizons, does the presence of timelike singularities and multiple copies of the asymptotic boundaries lead to new “bouncing” saddles in addition to those we have considered (see also \cite{Ceplak:2025dds, Dodelson:2025jff, AliAhmad:2026wem})? What happens in dimension $d\neq 4$? Are there connections with the Heun equation and integrability, as in \cite{Dodelson:2022yvn}?
What are the broader implications of our results for the physics of the black hole interior and ways to probe it, including quasinormal modes \cite{Motl:2003cd}, near-singularity Kasner behavior \cite{Frenkel:2020ysx, DeClerck:2025mem} and stringy features \cite{Martinec:1994xj, Zigdon:2024ljl}?

\section*{Acknowledgments}
We thank Nejc \v{C}eplak, Sean Hartnoll,  Yoav Zigdon and especially Keivan Namjou for discussions. The work of N.A.-J., S.C.-H. and J.-C. is supported by the National Science and Engineering Council of Canada (NSERC), funding reference SAPIN/00028-2022, and the Canada Research Chair program, reference number CRC-2022-00421.  The work of A.M. is supported in part by the Simons Foundation Grant No. 12574 and the Natural Sciences and Engineering Research Council of
Canada (NSERC), funding reference number SAPIN/00047-2020.

\bibliographystyle{apsrev4-1}
\bibliography{citation.bib}

@article{Hatta:2008tx,
    author = "Hatta, Y. and Iancu, E. and Mueller, A. H.",
    title = "{Jet evolution in the N=4 SYM plasma at strong coupling}",
    eprint = "0803.2481",
    archivePrefix = "arXiv",
    primaryClass = "hep-th",
    doi = "10.1088/1126-6708/2008/05/037",
    journal = "JHEP",
    volume = "05",
    pages = "037",
    year = "2008"
}

@article{Policastro:2002se,
    author = "Policastro, Giuseppe and Son, Dam T. and Starinets, Andrei O.",
    title = "{From AdS / CFT correspondence to hydrodynamics}",
    eprint = "hep-th/0205052",
    archivePrefix = "arXiv",
    reportNumber = "INT-PUB-02-32",
    doi = "10.1088/1126-6708/2002/09/043",
    journal = "JHEP",
    volume = "09",
    pages = "043",
    year = "2002"
}

@article{Policastro:2001yc,
    author = "Policastro, G. and Son, Dan T. and Starinets, Andrei O.",
    title = "{The Shear viscosity of strongly coupled N=4 supersymmetric Yang-Mills plasma}",
    eprint = "hep-th/0104066",
    archivePrefix = "arXiv",
    reportNumber = "NYU-TH-01-04-02, SNS-PH-01-05",
    doi = "10.1103/PhysRevLett.87.081601",
    journal = "Phys. Rev. Lett.",
    volume = "87",
    pages = "081601",
    year = "2001"
}

@article{Teaney:2006nc,
    author = "Teaney, Derek",
    title = "{Finite temperature spectral densities of momentum and R-charge correlators in N=4 Yang Mills theory}",
    eprint = "hep-ph/0602044",
    archivePrefix = "arXiv",
    doi = "10.1103/PhysRevD.74.045025",
    journal = "Phys. Rev. D",
    volume = "74",
    pages = "045025",
    year = "2006"
}

@article{Fidkowski:2003nf,
    author = "Fidkowski, Lukasz and Hubeny, Veronika and Kleban, Matthew and Shenker, Stephen",
    title = "{The Black hole singularity in AdS / CFT}",
    eprint = "hep-th/0306170",
    archivePrefix = "arXiv",
    reportNumber = "SU-ITP-03-16",
    doi = "10.1088/1126-6708/2004/02/014",
    journal = "JHEP",
    volume = "02",
    pages = "014",
    year = "2004"
}

@article{Festuccia:2005pi,
    author = "Festuccia, Guido and Liu, Hong",
    title = "{Excursions beyond the horizon: Black hole singularities in Yang-Mills theories. I.}",
    eprint = "hep-th/0506202",
    archivePrefix = "arXiv",
    reportNumber = "MIT-CTP-3641",
    doi = "10.1088/1126-6708/2006/04/044",
    journal = "JHEP",
    volume = "04",
    pages = "044",
    year = "2006"
}

@article{Ceplak:2024bja,
    author = "\v{C}eplak, Nejc and Liu, Hong and Parnachev, Andrei and Valach, Samuel",
    title = "{Black Hole Singularity from OPE}",
    eprint = "2404.17286",
    archivePrefix = "arXiv",
    primaryClass = "hep-th",
    month = "4",
    year = "2024"
}

@article{Basha:2018bvi,
    author = "Basha, Roi and Itzhaki, Nissan and Liram, Lior",
    title = "{High Energy Reflection Coefficients of Black Holes and Branes}",
    eprint = "1808.02036",
    archivePrefix = "arXiv",
    primaryClass = "hep-th",
    month = "8",
    year = "2018"
}

@article{Myers:2007we,
    author = "Myers, Robert C. and Starinets, Andrei O. and Thomson, Rowan M.",
    title = "{Holographic spectral functions and diffusion constants for fundamental matter}",
    eprint = "0706.0162",
    archivePrefix = "arXiv",
    primaryClass = "hep-th",
    doi = "10.1088/1126-6708/2007/11/091",
    journal = "JHEP",
    volume = "11",
    pages = "091",
    year = "2007"
}

@article{Buric:2025anb,
    author = "Buri\'c, Ilija and Gusev, Ivan and Parnachev, Andrei",
    title = "{Thermal holographic correlators and KMS condition}",
    eprint = "2505.10277",
    archivePrefix = "arXiv",
    primaryClass = "hep-th",
    month = "6",
    year = "2025"
}

@article{Caron-Huot:2025hmk,
    author = "Caron-Huot, Simon and Chakravarty, Joydeep and Namjou, Keivan",
    title = "{Looking at bulk points in general geometries}",
    eprint = "2502.14963",
    archivePrefix = "arXiv",
    primaryClass = "hep-th",
    doi = "10.1007/JHEP06(2025)197",
    journal = "JHEP",
    volume = "06",
    pages = "197",
    year = "2025"
}

@article{DeClerck:2025mem,
    author = "De Clerck, Marine and Hartnoll, Sean A. and Yang, Ming",
    title = "{Wheeler-DeWitt wavefunctions for 5d BKL dynamics, automorphic L-functions and complex primon gases}",
    eprint = "2507.08788",
    archivePrefix = "arXiv",
    primaryClass = "hep-th",
    month = "7",
    year = "2025"
}

@article{Klosch:1995qv,
    author = "Klosch, Thomas and Strobl, Thomas",
    title = "{Classical and quantum gravity in (1+1)-dimensions. Part 2: The Universal coverings}",
    eprint = "gr-qc/9511081",
    archivePrefix = "arXiv",
    reportNumber = "TUW-95-23, PITHA-95-24",
    doi = "10.1088/0264-9381/13/9/007",
    journal = "Class. Quant. Grav.",
    volume = "13",
    pages = "2395--2422",
    year = "1996"
}

@article{Caron-Huot:2025she,
    author = "Caron-Huot, Simon and Chakravarty, Joydeep and Namjou, Keivan",
    title = "{Boundary imprint of bulk causality}",
    eprint = "2501.13182",
    archivePrefix = "arXiv",
    primaryClass = "hep-th",
    month = "1",
    year = "2025"
}

@article{Amado:2008hw,
    author = "Amado, Irene and Hoyos-Badajoz, Carlos",
    title = "{AdS black holes as reflecting cavities}",
    eprint = "0807.2337",
    archivePrefix = "arXiv",
    primaryClass = "hep-th",
    reportNumber = "IFT-UAM-CSIC-08-45",
    doi = "10.1088/1126-6708/2008/09/118",
    journal = "JHEP",
    volume = "09",
    pages = "118",
    year = "2008"
}

@article{Zigdon:2024ljl,
    author = "Zigdon, Yoav",
    title = "{Stringy forces in the black hole interior}",
    eprint = "2407.12903",
    archivePrefix = "arXiv",
    primaryClass = "hep-th",
    doi = "10.1007/JHEP11(2024)063",
    journal = "JHEP",
    volume = "11",
    pages = "063",
    year = "2024"
}

@article{Maldacena:2001kr,
    author = "Maldacena, Juan Martin",
    title = "{Eternal black holes in anti-de Sitter}",
    eprint = "hep-th/0106112",
    archivePrefix = "arXiv",
    reportNumber = "NSF-ITP-01-59",
    doi = "10.1088/1126-6708/2003/04/021",
    journal = "JHEP",
    volume = "04",
    pages = "021",
    year = "2003"
}

@article{Festuccia:2008zx,
    author = "Festuccia, Guido and Liu, Hong",
    title = "{A Bohr-Sommerfeld quantization formula for quasinormal frequencies of AdS black holes}",
    eprint = "0811.1033",
    archivePrefix = "arXiv",
    primaryClass = "gr-qc",
    reportNumber = "MIT-CTP-3995, SCIPP-08-11",
    doi = "10.1166/asl.2009.1029",
    journal = "Adv. Sci. Lett.",
    volume = "2",
    pages = "221--235",
    year = "2009"
}

@article{Maldacena:1997re,
    author = "Maldacena, Juan Martin",
    title = "{The Large N limit of superconformal field theories and supergravity}",
    eprint = "hep-th/9711200",
    archivePrefix = "arXiv",
    reportNumber = "HUTP-97-A097, HUTP-98-A097",
    doi = "10.1023/A:1026654312961",
    journal = "Int. J. Theor. Phys.",
    volume = "38",
    pages = "1113--1133",
    year = "1999"
}

@article{Witten-ads-and-holography,
    author = "Witten, Edward",
    title = "{Anti-de Sitter space and holography}",
    eprint = "hep-th/9802150",
    archivePrefix = "arXiv",
    reportNumber = "IASSNS-HEP-98-15",
    doi = "10.4310/ATMP.1998.v2.n2.a2",
    journal = "Adv. Theor. Math. Phys.",
    volume = "2",
    pages = "253--291",
    year = "1998"
}

@article{Gubser:1998bc,
    author = "Gubser, S.S. and Klebanov, Igor R. and Polyakov, Alexander M.",
    title = "{Gauge theory correlators from noncritical string theory}",
    eprint = "hep-th/9802109",
    archivePrefix = "arXiv",
    reportNumber = "PUPT-1767",
    doi = "10.1016/S0370-2693(98)00377-3",
    journal = "Phys. Lett. B",
    volume = "428",
    pages = "105--114",
    year = "1998"
}

@article{Dodelson:2023nnr,
    author = "Dodelson, Matthew and Iossa, Cristoforo and Karlsson, Robin and Lupsasca, Alexandru and Zhiboedov, Alexander",
    title = "{Black hole bulk-cone singularities}",
    eprint = "2310.15236",
    archivePrefix = "arXiv",
    primaryClass = "hep-th",
    reportNumber = "CERN-TH-2023-192",
    doi = "10.1007/JHEP07(2024)046",
    journal = "JHEP",
    volume = "07",
    pages = "046",
    year = "2024"
}

@article{Grinberg:2020fdj,
    author = "Grinberg, Matan and Maldacena, Juan",
    title = "{Proper time to the black hole singularity from thermal one-point functions}",
    eprint = "2011.01004",
    archivePrefix = "arXiv",
    primaryClass = "hep-th",
    doi = "10.1007/JHEP03(2021)131",
    journal = "JHEP",
    volume = "03",
    pages = "131",
    year = "2021"
}

@article{Martinec:1994xj,
    author = "Martinec, Emil J.",
    title = "{Space - like singularities and string theory}",
    eprint = "hep-th/9412074",
    archivePrefix = "arXiv",
    reportNumber = "EFI-94-62",
    doi = "10.1088/0264-9381/12/4/005",
    journal = "Class. Quant. Grav.",
    volume = "12",
    pages = "941--950",
    year = "1995"
}

@article{Dodelson:2023vrw,
    author = "Dodelson, Matthew and Iossa, Cristoforo and Karlsson, Robin and Zhiboedov, Alexander",
    title = "{A thermal product formula}",
    eprint = "2304.12339",
    archivePrefix = "arXiv",
    primaryClass = "hep-th",
    reportNumber = "CERN-TH-2023-062",
    doi = "10.1007/JHEP01(2024)036",
    journal = "JHEP",
    volume = "01",
    pages = "036",
    year = "2024"
}

@article{Barrat:2025nvu,
    author = "Barrat, Julien and Bozkurt, Deniz N. and Marchetto, Enrico and Miscioscia, Alessio and Pomoni, Elli",
    title = "{The analytic bootstrap at finite temperature}",
    eprint = "2506.06422",
    archivePrefix = "arXiv",
    primaryClass = "hep-th",
    reportNumber = "DESY-25-078",
    month = "6",
    year = "2025"
}

@article{Motl:2003cd,
    author = "Motl, Lubos and Neitzke, Andrew",
    title = "{Asymptotic black hole quasinormal frequencies}",
    eprint = "hep-th/0301173",
    archivePrefix = "arXiv",
    reportNumber = "HEP-UK-0017, HUTP-03-A005",
    doi = "10.4310/ATMP.2003.v7.n2.a4",
    journal = "Adv. Theor. Math. Phys.",
    volume = "7",
    number = "2",
    pages = "307--330",
    year = "2003"
}

@article{Buric:2025fye,
    author = "Buri{\'c}, Ilija and Gusev, Ivan and Parnachev, Andrei",
    title = "{Holographic Correlators from Thermal Bootstrap}",
    eprint = "2508.08373",
    archivePrefix = "arXiv",
    primaryClass = "hep-th",
    month = "8",
    year = "2025"
}

@article{Bu:2020jfo,
    author = "Bu, Yanyan and Demircik, Tuna and Lublinsky, Michael",
    title = "{All order effective action for charge diffusion from Schwinger-Keldysh holography}",
    eprint = "2012.08362",
    archivePrefix = "arXiv",
    primaryClass = "hep-th",
    doi = "10.1007/JHEP05(2021)187",
    journal = "JHEP",
    volume = "05",
    pages = "187",
    year = "2021"
}

@article{Frenkel:2020ysx,
    author = "Frenkel, Alexander and Hartnoll, Sean A. and Kruthoff, Jorrit and Shi, Zhengyan D.",
    title = "{Holographic flows from CFT to the Kasner universe}",
    eprint = "2004.01192",
    archivePrefix = "arXiv",
    primaryClass = "hep-th",
    doi = "10.1007/JHEP08(2020)003",
    journal = "JHEP",
    volume = "08",
    pages = "003",
    year = "2020"
}

@article{Kontsevich:2021dmb,
    author = "Kontsevich, Maxim and Segal, Graeme",
    title = "{Wick Rotation and the Positivity of Energy in Quantum Field Theory}",
    eprint = "2105.10161",
    archivePrefix = "arXiv",
    primaryClass = "hep-th",
    doi = "10.1093/qmath/haab027",
    journal = "Quart. J. Math. Oxford Ser.",
    volume = "72",
    number = "1-2",
    pages = "673--699",
    year = "2021"
}

@article{AliAhmad:2026wem,
    author = "Ali Ahmad, Shadi and Almheiri, Ahmed and Lin, Simon",
    title = "{Continuing past the inner horizon using WKB}",
    eprint = "2601.02354",
    archivePrefix = "arXiv",
    primaryClass = "hep-th",
    month = "1",
    year = "2026"
}

@article{Dodelson:2025jff,
    author = "Dodelson, Matthew and Iossa, Cristoforo and Karlsson, Robin",
    title = "{Bouncing off a stringy singularity}",
    eprint = "2511.09616",
    archivePrefix = "arXiv",
    primaryClass = "hep-th",
    month = "11",
    year = "2025"
}

@article{Ceplak:2025dds,
    author = "{\v{C}}eplak, Nejc and Liu, Hong and Parnachev, Andrei and Valach, Samuel",
    title = "{Fooling the Censor: Going beyond inner horizons with the OPE}",
    eprint = "2511.09638",
    archivePrefix = "arXiv",
    primaryClass = "hep-th",
    month = "11",
    year = "2025"
}

@article{Witten:2021nzp,
    author = "Witten, Edward",
    title = "{A Note On Complex Spacetime Metrics}",
    eprint = "2111.06514",
    archivePrefix = "arXiv",
    primaryClass = "hep-th",
    month = "11",
    year = "2021"
}

@article{Son:2002sd,
    author = "Son, Dam T. and Starinets, Andrei O.",
    title = "{Minkowski space correlators in AdS / CFT correspondence: Recipe and applications}",
    eprint = "hep-th/0205051",
    archivePrefix = "arXiv",
    reportNumber = "INT-PUB-02-34",
    doi = "10.1088/1126-6708/2002/09/042",
    journal = "JHEP",
    volume = "09",
    pages = "042",
    year = "2002"
}

@article{Chakrabarty:2019aeu,
    author = "Chakrabarty, Bidisha and Chakravarty, Joydeep and Chaudhuri, Soumyadeep and Jana, Chandan and Loganayagam, R. and Sivakumar, Akhil",
    title = "{Nonlinear Langevin dynamics via holography}",
    eprint = "1906.07762",
    archivePrefix = "arXiv",
    primaryClass = "hep-th",
    doi = "10.1007/JHEP01(2020)165",
    journal = "JHEP",
    volume = "01",
    pages = "165",
    year = "2020"
}

@article{Glorioso:2018mmw,
    author = "Glorioso, Paolo and Crossley, Michael and Liu, Hong",
    title = "{A prescription for holographic Schwinger-Keldysh contour in non-equilibrium systems}",
    eprint = "1812.08785",
    archivePrefix = "arXiv",
    primaryClass = "hep-th",
    reportNumber = "MIT-CTP/5095; EFI-18-21",
    month = "12",
    year = "2018"
}

@article{Loganayagam:2022zmq,
    author = "Loganayagam, R. and Rangamani, Mukund and Virrueta, Julio",
    title = "{Holographic open quantum systems: toy models and analytic properties of thermal correlators}",
    eprint = "2211.07683",
    archivePrefix = "arXiv",
    primaryClass = "hep-th",
    doi = "10.1007/JHEP03(2023)153",
    journal = "JHEP",
    volume = "03",
    pages = "153",
    year = "2023"
}

@article{Barrat:2025twb,
    author = "Barrat, Julien and Bozkurt, Deniz N. and Marchetto, Enrico and Miscioscia, Alessio and Pomoni, Elli",
    title = "{Analytic thermal bootstrap meets holography}",
    eprint = "2510.20894",
    archivePrefix = "arXiv",
    primaryClass = "hep-th",
    reportNumber = "DESY-25-139 , YITP-SB-2025-16",
    month = "10",
    year = "2025"
}

@article{Chakravarty:2025ncy,
    author = "Chakravarty, Joydeep",
    title = "{Imprint of the black hole interior on thermal four-point correlators}",
    eprint = "2512.10912",
    archivePrefix = "arXiv",
    primaryClass = "hep-th",
    month = "12",
    year = "2025"
}

@article{Dodelson:2022yvn,
    author = "Dodelson, Matthew and Grassi, Alba and Iossa, Cristoforo and Panea Lichtig, Daniel and Zhiboedov, Alexander",
    title = "{Holographic thermal correlators from supersymmetric instantons}",
    eprint = "2206.07720",
    archivePrefix = "arXiv",
    primaryClass = "hep-th",
    reportNumber = "CERN-TH-2022-095",
    doi = "10.21468/SciPostPhys.14.5.116",
    journal = "SciPost Phys.",
    volume = "14",
    number = "5",
    pages = "116",
    year = "2023"
}

@article{Iliesiu:2018fao,
    author = "Iliesiu, Luca and Kolo{\u{g}}lu, Murat and Mahajan, Raghu and Perlmutter, Eric and Simmons-Duffin, David",
    title = "{The Conformal Bootstrap at Finite Temperature}",
    eprint = "1802.10266",
    archivePrefix = "arXiv",
    primaryClass = "hep-th",
    reportNumber = "CALT-TH-2018-013, PUPT-2550",
    doi = "10.1007/JHEP10(2018)070",
    journal = "JHEP",
    volume = "10",
    pages = "070",
    year = "2018"
}

@article{Fitzpatrick:2019zqz,
    author = "Fitzpatrick, A. Liam and Huang, Kuo-Wei",
    title = "{Universal Lowest-Twist in CFTs from Holography}",
    eprint = "1903.05306",
    archivePrefix = "arXiv",
    primaryClass = "hep-th",
    doi = "10.1007/JHEP08(2019)138",
    journal = "JHEP",
    volume = "08",
    pages = "138",
    year = "2019"
}

@article{Caron-Huot:2009ypo,
    author = "Caron-Huot, S.",
    title = "{Asymptotics of thermal spectral functions}",
    eprint = "0903.3958",
    archivePrefix = "arXiv",
    primaryClass = "hep-ph",
    doi = "10.1103/PhysRevD.79.125009",
    journal = "Phys. Rev. D",
    volume = "79",
    pages = "125009",
    year = "2009"
}
\appendix
\section{High-frequency expansion of $G_{\rm ret}(\omega)$} \label{app:expansions}

\def\tphi{\chi}

Here we present our algorithm to expand the retarded correlator at large $(\omega^2-k^2)$ when $d$ is an even integer. The algorithm can be viewed as a momentum space version of that in \cite{Fitzpatrick:2019zqz}. The perturbative contributions originate from the region where $r$ is large but
the ratio $z=\ell^2\frac{\sqrt{\omega^2-k^2}}{r}$ is fixed. In the limit, the wave equation \eqref{eom} expressed in terms of $\phi(r)=((r/r_h)^d-1)^{-1/2}\tphi(z)$ becomes the Bessel equation plus perturbative corrections:
\begin{equation} \label{Bessel plus corrections}
\left[z^2\partial_z^2 + z \partial_z +z^2-\nu^2\right]\tphi(z) \equiv {\cal D}_\nu \tphi(z) = U\tphi(z),    
\end{equation}
where $\nu=\Delta-\frac{d}{2}$. The perturbing potential is
\begin{align}
    U&=\epsilon z^d\left[\frac{\nu^2-z^2}{1-\epsilon z^d}-\frac{\gamma^2z^2+(d/2)^2}{(1-\epsilon z^d)^2}\right]
    \nonumber\\
&=\sum_{n=1}^\infty \epsilon^n z^{nd}
\Bigl[ \nu^2-z^2-n\bigl(\gamma^2z^2+(d/2)^2\bigr)\Bigr]
\label{Upot}
\end{align}
where $\epsilon=\left(\frac{4\pi T/d}{\sqrt{\omega^2-q^2}}\right)^d$ is a dimensionless ratio of the energy density and probe energy and $\gamma=\omega/\sqrt{\omega^2-q^2}$.

At order $\epsilon^0$, the solution to \eqref{Bessel plus corrections} are Bessel/Hankel functions. Infalling boundary conditions at the horizon pick the solution with asymptotic $\tphi\propto e^{\ic z}$
(up to nonperturbative effects that do not contribute at any finite order in $1/\omega$):
\begin{equation}
    {\cal D}_\nu\tphi^{(0)}(z)=0\quad\Rightarrow\quad \tphi^{(0)}(z)\propto H^{(1)}_\nu(z).
\end{equation}
We see from \eqref{Upot} that to calculate the ${\cal O}(\epsilon^1)$ correction to $\tphi$, we need to solve the inhomogeneous equation
${\cal D}_\nu \delta\tphi=z^a H^{(1)}_\nu$ with $a\in \{d,d+2\}$.
In general this cannot be solved in terms of elementary functions and therefore the procedure becomes difficult to iterate. However, when $d$ is an even integer, elementary solutions turn out to exist.

Our method consists in making an ansatz for $\phi(z)$ in terms of polynomials multiplying Bessel functions. This was inspired by the position space ansatz in \cite{Fitzpatrick:2019zqz}, which used terms of the form $(t^2-\vec{x^2})^a t^b$; the Fourier transform of each term gives a Bessel function times a polynomial when $b$ is an even integer (this is the only case where the procedure works).
Staying in momentum space, we find that it suffices to use the original Bessel function, and one with index shifted by 1, for example:
\begin{subequations}\label{Dinverse}
\begin{align}\label{simplest Dinverse}
{\cal D}_\nu^{-1}\left(2z^2 H_\nu^{(1)} \right)
&=z H_{\nu+1}^{(1)},\\
\label{simplest Dinverse 2}
{\cal D}_\nu^{-1}\left(4 z^3 H_{\nu+1}^{(1)}\right)&=
-z^2 H_{\nu}^{(1)}+2z(1{+}\nu)H_{\nu+1}^{(1)},
\\
{\cal D}_\nu^{-1}\left(6z^4 H_\nu^{(1)} \right)
&=z^2(1{-}\nu) H_{\nu}^{(1)}+
z(z^2{+}2\nu^2{-}2) H_{\nu+1}^{(1)}.
\end{align}
\end{subequations}
Here the argument of every Hankel function is $z$.
The second line illustrates that the procedure can be iterated: the set of functions $H_{\nu}$ and $H_{\nu+1}$ times polynomials form a closed set under the action of ${\cal D}_{\nu}^{-1}$.
This enables solving \eqref{Bessel plus corrections} to arbitrarily high order, at least when $d$ is an even integer.
The solutions \eqref{Dinverse} are unchanged if $H_c^{(1)}$ is replaced by any other Bessel function (e.g. $H_c^{(2)}$ or $J_c$ or $Y_c$).

The inverse ${\cal D}_\nu^{-1}$ is unique up to homogeneous solutions; we impose the (perturbative) infalling condition and vanishing of the $z^{-\nu}$ coefficient, which modifies \eqref{Dinverse} by addition of $H_\nu^{(1)}$ terms.
The following recursion then efficiently computes all required primitives:
\begin{equation}\label{Dinverse recursion}\begin{aligned}
{\cal D}_\nu^{-1}(z^k H_\nu^{(1)})
&=
\frac{z^{k-1}H_{\nu+1}^{(1)}}{2(k-1)}-
\frac{n_{k,\nu} %(k{-}2)(k{-}2{-}2\nu)
}{2(k-1)}
{\cal D}_\nu^{-1}(z^{k-1} H_{\nu+1}),
\\
{\cal D}_\nu^{-1}(z^k H_{\nu+1}^{(1)})
&=-\frac{z^{k-1}H_\nu^{(1)}}{2(k-1)}
+\frac{k{-}1{+}2\nu}{2}{\cal D}_\nu^{-1}(z^{k-1} H_{\nu}^{(1)})
\end{aligned}\end{equation}
with $n_{k,\nu}=(k-2)(k-2-2\nu)$. We use these for $k\geq 4$ even in the first line and $k\geq 3$ odd in the second line. The recursion then terminates on (the modified) \eqref{simplest Dinverse}.
    
Using this method, for example, for $\Delta=4$ (equivalent to the shear channel stress-tensor correlator $G_R^{xy,xy}$), we find the following order-by-order solution to the wave equation
\begin{align}
       & \tphi^{\Delta=4}_{\gamma=1}(z)=
    \left[1-\epsilon^2\left(\frac{2z^6}{15}+\frac{z^8}{45}+\frac{z^{10}}{50}\right)+{\cal O}(\epsilon^3)\right]H_{2}(z)
\nonumber\\\quad &+
\left[-\epsilon\frac{z^5}{5} +\epsilon^2\left(\frac{4z^5}{5}-\frac{2z^7}{15}-\frac{z^{9}}{90}\right)+{\cal O}(\epsilon^3)\right]H_{3}(z)  .
\end{align}
It is also straightforward to retain the $\gamma$ dependence, which is polynomial, and so we record it below. Taking the ratio of the $z^{\nu}$ and $z^{-\nu}$ coefficients as $z\to 0$
according to the standard recipe \eqref{recipe}, we thus obtain
the retarded Green's function:
\begin{align} \label{GR exp}
    &G_{\rm ret}^{\Delta=4}(\omega,q) \propto
(\omega^2{-}q^2)^2\Bigg[
{-}\log\frac{q^2{-}(\omega{+}\ic 0)^2}{\mu^2}
\nonumber\\ &\quad -\frac{64\epsilon}{5}(\gamma^2+1)+\frac{512\epsilon^2}{35}(28\gamma^4-24\gamma^2+3)
+{\cal O}(\epsilon^3)
\Bigg].
\end{align}
The $\log$ term is the $T=0$ result, which
also controls the short-distance limit of the correlator;
the overall convention-dependent normalization, left unspecified here, could be determined from this limit. Note that the $T=0$ term is logarithmically ultraviolet divergent, whence the arbitrary holographic renormalization scale $\mu$; this generically happens for integer $\Delta$. For real positive $\omega$, the imaginary part of the square bracket, $\ic\pi$, comes entirely from the first term as explained in the text.  We expect the expansion \eqref{GR exp} to be valid for spacelike or timelike momentum ($\gamma<1$ or $\gamma>1$), although its validity near the lightcone [when $\epsilon\gamma^2\propto T^4\omega^2/(\omega^2-q^2)^3\gtrsim 1$] is less clear (similar scales appeared in \cite{Hatta:2008tx}).

The method just detailed enables the calculation of $1/\omega$ series to very high order in $\epsilon$. The bottleneck is multiplication of the potential $U$ by the lower-order solutions, and multiplication of the resulting coefficients with the tabulated ${\cal D}_{\nu}^{-1}$ primitives (which are obtained essentially instantaneously from \eqref{Dinverse recursion}).
%For a fixed numerical value of $\nu$, the primitives \eqref{Dinverse recursion} can be tabulated up to $k\sim$ thousand in a few seconds on a laptop.
Using exact rational arithmetic in our {\it Mathematica} implementation, obtaining the correlator to order $\epsilon^{400}$ for $\Delta=4$ and $q=0$ takes about an hour on a laptop.

\section{Asymptotics in difference spaces} \label{app:Fourier}

Here we record various formulas relating the small-time and large-frequency expansions of the Wightman functions, focusing on zero spatial momentum $(q=0)$. We reproduce here for convenience the expansion coefficients defined in the text for the retarded function in \eqref{Gret t} and \eqref{Gret omega}:
\begin{align} \label{Gret t app}
    G_{\rm ret}(t,q{=}0)&= t^{-1-2\nu}
    \sum_{n=0}^\infty a_n \left(\frac{t}{\beta}\right)^{nd}\qquad(t>0) ,
\\ \label{Gret omega app}
    G_{\rm ret}(\omega,q{=}0) &\sim (-\ic\omega)^{2\nu}
\sum_{n=0}^\infty \frac{a_n}{(-\ic \beta\omega)^{nd}} 
\Gamma(nd-2\nu)
\end{align}
where we recall that $\nu=\Delta_{\cal O}-\frac{d}{2}$.
Taking the imaginary part of the second gives the asymptotic series for $G^>(\omega,q{=}0)=\frac{2{\rm Im} G_{\rm ret}(\omega)}{1-e^{-\beta\omega}}$, which is the same, up to corrections $\sim e^{-\beta\omega}$ that are nonperturbative at large $\omega$, as that for $2{\rm Im} G_{\rm ret}(\omega)$:
\begin{equation} \label{G> t omega}
     G^>(\omega)\sim \sum_{n=0}^\infty \frac{a_n \omega^{2\nu-nd}\beta^{-nd}}{\Gamma(1+2\nu-nd)
     \cos(\frac{\pi}2(2\nu-n d))} .
\end{equation}
Fourier-transforming back to the time domain, this high-frequency expansion only determines the \emph{nonanalytic} terms in $G^>(t)$ near $t=0$:
\begin{align} \label{G> t app}
    G^>(t,q{=}0) &=  (\ic t)^{-1-2\nu}
    \sum_{n=0}^\infty \frac{a_n}{\cos(\frac{\pi}2(n d-2\nu))} \left(\frac{\ic t}{\beta}\right)^{nd} \nonumber\\ &\quad+
    \mbox{(integer powers of $t^2$)}.
\end{align}
The second line can be interpreted as the contribution to the OPE from ${\cal O}{\cal O}$ double traces.  Note that we have written $G^>(t)$ in a form which respects the analyticity of this correlator in the strip $-\beta<{\rm Im} t<0$, i.e. a small negative imaginary part is allowed.

Using the complex conjugate expansion for $G^<(t)$, \eqref{Gret t app} can be verified directly to be the discontinuity of \eqref{G> t app}, that is, $G_{\rm ret}=\ic(G^>-G^<)$.

Starting from the large-frequency expansion of $G_{\rm ret}(\omega)$ in \eqref{Gret omega app}, the coefficients $a_n$ can thus be obtained by dividing each term by a Gamma function and a phase.

Although closely related expansions have been considered in the literature, direct comparison did not seem possible since results typically are for $\vec{x}=0$
[for example (3.9) of \cite{Ceplak:2024bja}] whereas we focus on $\vec{q}=0$.
The full Gegenbauer expansion in
\eqref{OPE general} (for the retarded function) could be obtained in future work by Fourier-transforming  \eqref{GR exp} termwise, keeping all the dependence on $\gamma$.

Consider now a nonperturbative contribution $G_{\rm ret}(\omega,q=0)\supset  b_p e^{\ic \omega t_*} \omega^p$ appearing in a large-$\omega$ transseries such as \eqref{Gret transseries}.
Evaluating the Fourier transform  on the contour in Fig.~\ref{fig:contour}, this is equivalent to a $t$-plane singularity:
\begin{equation} \label{Gt asympt}
G_{\rm ret}(\omega)\!\supset\! b_p \omega^p e^{\ic\omega t_*} \Leftrightarrow 
    G_{\rm ret}(t)\!\supset\! \frac{b_p \Gamma(p+1)}{2\pi}\frac{1}{(\ic(t{-}t_*))^{p+1}} .
\end{equation}
Let us translate this into a prediction for the large-order behavior of the coefficients $a_n$ around $t=0$, assuming that the singularity at $t_*$ is the one closest to the origin.
%We will also assume that $t_*$ lies in the analyticity strip of $G^<$, that is $0<{\rm Im} t_*<\beta$, so that $G_{\rm ret}$ and $\ic G^>$ contain the same singularity.
The basic idea is to use the following identity for the singular behavior as $x\to 1$ of a sum with power-law coefficients:
\begin{align} \label{large x identity}
    \sum_{n=1}^\infty n^p x^n &= \frac{\Gamma(p+1)}{(1{-}x)^{p+1}} \left(
    1-\frac{p+1}{2} (1{-}x)+\mathcal{O}((1{-}x)^2)\right) 
    \nonumber\\ &\quad+\mbox{(terms analytic at $x\to 1$).}
\end{align}
Applying this to $x=(t/t_*)^d$ and comparing with \eqref{Gret t app},
\begin{equation} \label{largen lim}
    \lim_{n\to \infty}a_n = n^p\frac{b_p (\ic d)^{p+1}}{2\pi  t_*^{p-2\nu}}\left(\frac{\beta}{t_*}\right)^{nd}\times r_{p,\nu}(n)
\end{equation}
with remainder $r_{p,\nu}(n)=1+\mathcal{O}(n^{-1})$.

The subleading corrections to \eqref{largen lim}, assuming a pure power law in \eqref{Gt asympt} the $\omega$ or $t$ planes,
are interesting because they can be detected numerically.
They have two origins. First, the pure power in $(t-t_*)$ in \eqref{Gt asympt} does not map to a pure power of $(1-x)$. Second, a pure power of $(1-x)$ does not translate into a pure power of $n$, due to the series of corrections in \eqref{large x identity}.
Both effects can be systematically accounted for order by order in $1/n$.
Accounting for these we obtain the more refined large-order prediction
\begin{equation}\begin{aligned} \label{r expansion}
    r_{p,\nu}(n) &= 1 + \frac{p}{n}\frac{p{-}1{-}4\nu}{2d}+\frac{p(p{-}1)}{n^2}
\\ &\hspace{-5mm}\times\frac{3p^2-p(24\nu{+}7)+48\nu^2{+}24\nu{+}2}{24 d^2}+ \mathcal{O}(n^{-3}) .
\end{aligned}\end{equation}

Applying these formulas to the transseries in \eqref{Gret transseries} including the subleading reflection coefficient in \eqref{R NLO}, we obtain the following prediction for the large-$n$ behavior of the expansion coefficients [in the same normalization as \eqref{Gret transseries}, where $a_0=\Gamma(2\nu+1)\cos(\pi\nu)/\pi$]:
\begin{equation}\begin{aligned} \label{large n app}
     \lim_{n\to\infty}\frac{a_n}{(-4)^{n}} &=
     (4n)^{2\nu} \frac{4}{\pi}r_{2\nu,\nu}
\\\ &\hspace{-14mm}+
(4n)^{2\nu-4/3}(2\pi)^{1/3} R^{(1)}r_{2\nu-4/3,\nu} +\mathcal{O}(n^{2\nu-8/3}) .
\end{aligned}\end{equation}

\section{Borel resummation for R-current correlators} \label{Borel R}

In this appendix we confirm the identification of the square bracket of \eqref{Gxx exact 2} with the canonical Borel resummation of the large-$\omega$ series.
Using the standard series expansion of $\psi^{(0)}$ we find
\begin{equation} \label{Gxx series}
    G_{\rm ret}^{xx}(\omega,q{=}0)\sim
\omega^2  \left[\ic\pi-2\log(\omega)
+\sum_{n=1}^\infty \left(\frac{-4}{\omega^4}\right)^n \frac{B_{4n}}{2n}\right]
\end{equation}
where $B_{n}$ are Bernoulli numbers. The $\sim$ indicates that this is an asymptotic expansion: the coefficients grow factorially. In the notation of \eqref{Gret omega app}, this gives the coefficients $a_0^{xx}=-4$ and 
\begin{equation}
 a_n^{xx}=-\frac{(-64\pi^4)^nB_{4n}}{2n\Gamma(4n-2)}\qquad(n\geq 1)\,.
\end{equation}
(Note that the normalization here is $2\pi$ larger than that used for scalar correlators in the rest of this paper.)
In the time domain the series \eqref{Gret t app} has a finite radius of convergence and, in fact, can be summed analytically:
\begin{equation}
    G_{\rm ret}^{xx}(t,q{=}0) = -\partial_t^2
\left[\frac{\pi(1-\ic)/\beta}{\tan\frac{\pi t}{\beta}(1-\ic)}+
\frac{\pi(1+\ic)/\beta}{\tan\frac{\pi t}{\beta}(1+\ic)}\right].
\end{equation}
The canonical Borel resummation of the series \eqref{Gxx series}, for large real $\omega$, is defined by integrating the Fourier transform over imaginary $t$ which is the steepest-descent path for $e^{\ic t \omega}$.
As depicted in Fig.~\ref{fig:contour}, the full Fourier transform over the real axis is equal to this, plus the sum over the poles at $t=k \frac{1+\ic}{2}\beta$.
Accounting for ultraviolet regularization and a small arc near the origin (responsible for the $\ic\pi$ below),  we can write these explicitly as
\begin{align}
&G_{\rm ret}^{xx}(\omega,q{=}0) \equiv \int_0^\infty \d t\, e^{\ic\omega t}G_{\rm ret}^{xx}(t,q{=}0)
\nonumber \\
&=\Biggl[
\int_0^\infty \d\tau\, e^{-\omega\tau}
\left(\ic G_{\rm ret}^{xx}(\ic\tau,q{=}0)-\frac{4}{\tau^3}\right)
\nonumber \\ &\quad
+\omega^2(\ic\pi-2\log(\omega)-2\gamma_E)\Bigg]
 + 2\pi \ic\, \omega^2 \sum_{k=1}^\infty e^{\ic (1+\ic)k\beta\omega/2}\,.
\end{align}
The square bracket and remainder can be verified to be precisely equal to those in \eqref{Gxx exact 2}. This confirms the bulk-boundary identification proposed in Sec.~\ref{sec:proposal}: canonically Borel-resumming the boundary correlator using the steepest-descent time contour, coincides with solving the bulk radial equation along a steepest-descent contour in $r$.

\end{document}